\documentclass[trackchanges]{aastex701}
\usepackage{gensymb}
\usepackage{caption}
\usepackage{subcaption}
\usepackage{amsmath}
\usepackage{multirow}
\usepackage[siunitx, RPvoltages]{circuitikz}
\usetikzlibrary{positioning, fit, backgrounds}

\begin{document}

\title{The Simons Observatory: Design and Initial Performance of the Detector Readout System for the Large-Aperture Telescope}

\author[orcid=0000-0002-6452-4220,gname='Thomas',sname='Satterthwaite']{Thomas P. Satterthwaite}
\affiliation{Department of Physics, Stanford University, 382 Via Pueblo Mall, Stanford, CA 94305, USA}
\affiliation{Kavli Institute for Particle Astrophysics and Cosmology, 452 Lomita Mall, Stanford, CA 94305, USA}
\email[show]{tpsatt@stanford.edu}

\author[orcid=0000-0002-9516-3245,gname='Tristan',sname='Pinsonneault-Marotte']{Tristan Pinsonneault-Marotte}
\affiliation{Kavli Institute for Particle Astrophysics and Cosmology, 452 Lomita Mall, Stanford, CA 94305, USA}
\affiliation{SLAC National Accelerator Laboratory, 2575 Sand Hill Road, Menlo Park, CA 94025, USA}
\email{tristpm@slac.stanford.edu}

\author[orcid=0000-0001-7878-4229,gname='Shawn',sname='Henderson']{Shawn W. Henderson}
\affiliation{Kavli Institute for Particle Astrophysics and Cosmology, 452 Lomita Mall, Stanford, CA 94305, USA}
\affiliation{SLAC National Accelerator Laboratory, 2575 Sand Hill Road, Menlo Park, CA 94025, USA}
\email{swh76@stanford.edu}

\author[orcid=0000-0002-9957-448X,gname='Zeeshan',sname='Ahmed']{Zeeshan Ahmed}
\affiliation{Kavli Institute for Particle Astrophysics and Cosmology, 452 Lomita Mall, Stanford, CA 94305, USA}
\affiliation{SLAC National Accelerator Laboratory, 2575 Sand Hill Road, Menlo Park, CA 94025, USA}
\email{zeesh@slac.stanford.edu}

\author[orcid=0000-0002-3407-5305,gname='Kam',sname='Arnold']{Kam Arnold}
\affiliation{Department of Astronomy and Astrophysics, University of California San Diego, 9500 Gilman Drive \#0424, La Jolla, CA 92093, USA}
\email{arnold@ucsd.edu}

\author[orcid=0009-0002-7336-6903,gname='Declan',sname='Baker']{Declan Baker}
\affiliation{Department of Physics and Astronomy, University of Pennsylvania, 209 South 33rd Street, Philadelphia, PA 19014, USA}
\email{decbaker@sas.upenn.edu}

\author[orcid=0000-0002-7888-6222,gname='Andrew',sname='Bazarko']{Andrew Bazarko}
\affiliation{Department of Physics, Princeton University, Jadwin Hall,  Washington Road, Princeton, NJ 08542, USA}
\email{bazarko@princeton.edu}

\author[orcid=0000-0002-1327-1921,gname='Josh',sname='Borrow']{Josh Borrow}
\affiliation{Department of Physics and Astronomy, University of Pennsylvania, 209 South 33rd Street, Philadelphia, PA 19014, USA}
\email{josh@joshborrow.com}

\author[orcid=0000-0002-0370-8077,gname='Michael',sname='Brown']{Michael L. Brown}
\affiliation{Jodrell Bank Centre for Astrophysics, Department of Physics and Astronomy, University of Manchester, Oxford Road, Manchester M13 9PL, UK}
\email{m.l.brown@manchester.ac.uk}

\author[orcid=0000-0002-3169-9761,gname='Mark',sname='Devlin']{Mark Devlin}
\affiliation{Department of Physics and Astronomy, University of Pennsylvania, 209 South 33rd Street, Philadelphia, PA 19014, USA}
\email{devlin@upenn.edu}

\author[orcid=0000-0002-1940-4289,gname='Simon',sname='Dicker']{Simon Dicker}
\affiliation{Department of Physics and Astronomy, University of Pennsylvania, 209 South 33rd Street, Philadelphia, PA 19014, USA}
\email{sdicker@physics.upenn.edu}

\author[orcid=0009-0006-8427-6259,gname='Peter',sname='Dow']{Peter N. Dow}
\affiliation{Department of Astronomy, University of Virginia, 530 McCormick Road, Charlottesville, VA 22904, USA}
\email{pd3cx@virginia.edu}

\author[orcid=0000-0002-9693-4478,gname='Shannon',sname='Duff']{Shannon M. Duff}
\affiliation{Quantum Sensors Division, National Institute of Standards and Technology, 325 Broadway, Boulder, CO 80305, USA}
\email{shannon.duff@nist.gov}

\author[orcid=0000-0002-9962-2058,gname='Daniel',sname='Dutcher']{Daniel Dutcher}
\affiliation{Department of Physics, Princeton University, Jadwin Hall,  Washington Road, Princeton, NJ 08542, USA}
\email{ddutcher@uchicago.edu}

\author[orcid=0000-0001-9880-3634,gname='John',sname='Groh']{John C. Groh}
\affiliation{Physics Division, Lawrence Berkeley National Laboratory, 1 Cyclotron Road, Berkeley, CA 94720, USA}
\email{john.groh@lbl.gov}

\author[orcid=0000-0001-6519-502X,gname='Saianeesh',sname='Haridas']{Saianeesh K. Haridas}
\affiliation{Department of Physics and Astronomy, University of Pennsylvania, 209 South 33rd Street, Philadelphia, PA 19014, USA}
\email{haridas@sas.upenn.edu}

\author[orcid=0000-0003-1248-9563,gname='Kathleen',sname='Harrington']{Kathleen Harrington}
\affiliation{High Energy Physics Division, Argonne National Laboratory, 9700 South Cass Avenue, Lemont, IL, 60439, USA}
\affiliation{Department of Astronomy and Astrophysics, University of Chicago, 5720 South Ellis Avenue, Chicago, IL, 60637, USA}
\email{kharrington@anl.gov}

\author[orcid=0000-0002-3757-4898,gname='Erin',sname='Healy']{Erin Healy}
\affiliation{Kavli Institute for Cosmological Physics, University of Chicago, 5640 South Ellis Avenue, Chicago, IL, 60637, USA}
\email{healye@uchicago.edu}

\author[orcid=0000-0002-2781-9302,gname='Johannes',sname='Hubmayr']{Johannes Hubmayr}
\affiliation{Quantum Sensors Division, National Institute of Standards and Technology, 325 Broadway, Boulder, CO 80305, USA}
\email{johannes.hubmayr@nist.gov}

\author[orcid=0000-0002-6898-8938,gname='Bradley',sname='Johnson']{Bradley R. Johnson}
\affiliation{Department of Astronomy, University of Virginia, 530 McCormick Road, Charlottesville, VA 22904, USA}
\email{bradley.johnson@virginia.edu}

\author[orcid=0000-0003-3118-5514,gname='Brian',sname='Keating']{Brian Keating}
\affiliation{Department of Physics, University of California San Diego, 9500 Gilman Drive \#0319, La Jolla, CA 92093, USA}
\email{bkeating@ucsd.edu}

\author[orcid=0000-0001-5374-1767,gname='Anna',sname='Kofman']{Anna M. Kofman}
\affiliation{Kavli Institute for Cosmological Physics, University of Chicago, 5640 South Ellis Avenue, Chicago, IL, 60637, USA}
\affiliation{Department of Physics, University of Chicago, 5720 South Ellis Avenue, Chicago, IL, 60637, USA}
\email{amkofman@uchicago.edu}

\author[orcid=0000-0003-3106-3218,gname='Adrian',sname='Lee']{Adrian T. Lee}
\affiliation{Department of Physics, University of California, Berkeley, 366 Physics North MC 7300, Berkeley, CA 94720, USA}
\affiliation{Physics Division, Lawrence Berkeley National Laboratory, 1 Cyclotron Road, Berkeley, CA 94720, USA}
\email{Adrian.Lee@berkeley.edu}

\author[orcid=0000-0003-4629-5759,gname='Alex',sname='Manduca']{Alex Manduca}
\affiliation{Department of Physics and Astronomy, University of Pennsylvania, 209 South 33rd Street, Philadelphia, PA 19014, USA}
\email{manduca@sas.upenn.edu}

\author[orcid=0009-0000-1028-3524,gname='Aashrita',sname='Mangu']{Aashrita Mangu}
\affiliation{Department of Physics, University of Chicago, 5720 South Ellis Avenue, Chicago, IL, 60637, USA}
\email{amangu@uchicago.edu}

\author[orcid=0000-0002-7340-9291,gname='Jenna',sname='Moore']{Jenna E. Moore}
\affiliation{Department of Physics, Duke University, 120 Science Drive, Durham, NC 27708, USA}
\email{jenna.moore@duke.edu}

\author[orcid=0000-0001-7125-3580,gname='Michael',sname='Niemack']{Michael D. Niemack}
\affiliation{Department of Physics, Cornell University, 142 Sciences Drive, Ithaca, NY 14853, USA}
\affiliation{Department of Astronomy, Cornell University, 122 Sciences Drive, Ithaca, NY 14853, USA}
\email{niemack@cornell.edu}

\author[orcid=0000-0003-2454-6828,gname='Sandra',sname='O'Neill']{Sandra O'Neill}
\affiliation{Department of Physics, Princeton University, Jadwin Hall,  Washington Road, Princeton, NJ 08542, USA}
\email{so3803@princeton.edu}

\author[orcid=0000-0003-1842-8104,gname='John',sname='Orlowski-Scherer']{John Orlowski-Scherer}
\affiliation{Department of Physics and Astronomy, University of Pennsylvania, 209 South 33rd Street, Philadelphia, PA 19014, USA}
\email{jorlo@sas.upenn.edu}

\author[orcid=0000-0001-5680-4989,gname='Yudai',sname='Seino']{Yudai Seino}
\affiliation{Department of Physics, Princeton University, Jadwin Hall,  Washington Road, Princeton, NJ 08542, USA}
\email{ys9136@princeton.edu}

\author[orcid=0000-0002-4561-7026,gname='Mufan',sname='Shao']{Mufan Shao}
\affiliation{Department of Physics, Princeton University, Jadwin Hall,  Washington Road, Princeton, NJ 08542, USA}
\email{ms1123@princeton.edu}

\author[orcid=0000-0002-9246-5571,gname='Carlos',sname='Sierra']{Carlos Sierra}
\affiliation{Kavli Institute for Particle Astrophysics and Cosmology, 452 Lomita Mall, Stanford, CA 94305, USA}
\affiliation{SLAC National Accelerator Laboratory, 2575 Sand Hill Road, Menlo Park, CA 94025, USA}
\email{csierra@stanford.edu}

\author[orcid=0000-0001-7480-4341,gname='Max',sname='Silva-Feaver']{Max Silva-Feaver}
\affiliation{Department of Physics, Yale University, 217 Prospect Street, New Haven, CT 06511, USA}
\affiliation{Wright Laboratory, Yale University, 272 Whitney Avenue, New Haven, CT 06511, USA}
\email{maximiliano.silva-feaver@yale.edu}

\author[orcid=0000-0002-7020-7301,gname='Suzanne',sname='Staggs']{Suzanne Staggs}
\affiliation{Department of Physics, Princeton University, Jadwin Hall,  Washington Road, Princeton, NJ 08542, USA}
\email{staggs@princeton.edu}

\author[orcid=0000-0002-2105-7589,gname='Eve',sname='Vavagiakis']{Eve M. Vavagiakis}
\affiliation{Department of Physics, Duke University, 120 Science Drive, Durham, NC 27708, USA}
\affiliation{Department of Physics, Cornell University, 142 Sciences Drive, Ithaca, NY 14853, USA}
\email{eve.vavagiakis@duke.edu}

\author[orcid=0000-0002-8710-0914,gname='Yuhan',sname='Wang']{Yuhan Wang}
\affiliation{Department of Physics, Cornell University, 142 Sciences Drive, Ithaca, NY 14853, USA}
\email{yw2684@cornell.edu}

\author[orcid=0000-0002-7567-4451,gname='Edward',sname='Wollack']{Edward J. Wollack}
\affiliation{NASA Goddard Space Flight Center, 8800 Greenbelt Road, Greenbelt, MD 20771, USA}
\email{edward.j.wollack@nasa.gov}

\correspondingauthor{Thomas P. Satterthwaite}

\begin{abstract}

We present the design, implementation, and initial performance of a highly multiplexed cryogenic superconducting detector readout system in the context of its deployment to the Simons Observatory large-aperture telescope (LAT). The Simons Observatory is a cosmic microwave background experiment located at 5,200\;m in Chile's Atacama Desert. In addition to its 6\;m LAT, whose broad range of science goals spans probing large-scale structure to searching for new Solar System objects, the observatory also currently includes three 0.42\;m small-aperture telescopes, which search for signatures of primordial gravitational waves. Nearly 100,000 optically-coupled transition-edge sensor bolometers have been deployed across the observatory's four telescopes to enable these science goals. The readout systems for these telescopes use microwave frequency multiplexing to simultaneously read out 860 optically-coupled bolometers per transmission line. This is enabled by the large-scale fabrication of multiplexing chips which house radio-frequency superconducting quantum interference devices for each detector, and the development of room-temperature electronics to read out these bolometers. This has enabled the largest deployment of superconducting sensors for astronomical observations to date. This paper focuses on the design and implementation of the readout system for the 63,000 detectors in the LAT's cryogenic receiver. We present performance results from an early phase in the telescope's operation. We show that the readout system's performance enables the instrument to meet key benchmarks, including successful readout of 81.4\% of its optically-coupled bolometers, which exceeds projections, and readout noise-equivalent current consistent with pre-deployment specifications.

\end{abstract}


\section{Introduction} \label{sec:intro}

The Simons Observatory is a cosmic microwave background (CMB) experiment located at 5,200\;m on Cerro Toco in the Atacama Desert of northern Chile. Thus far, four telescopes have been built at the observatory in order to enable sensitivity to a range of probes of fundamental physics \citep{SOScienceGoals-Ade_2019}. Three 0.42\;m refracting small-aperture telescopes (SATs) are mapping 10\% of the sky to search for the signatures of primordial gravitational waves. The 6\;m crossed-Dragone large-aperture telescope (LAT) is mapping 60\% of the sky with the aim of reaching $3\;\mu\text{K-arcmin}$ temperature map noise in CMB bands in order to probe the growth of large-scale structure, constrain the parameters of the $\Lambda$CDM model for concordance cosmology, and bound the sum of the neutrino masses, among other science goals \citep{ASO-Abitbol_2025,Galactic-Hensley_2022}. Overlap of the survey footprint of the LAT with that of complementary instruments observing in other regions of the electromagnetic spectrum, such as the Dark Energy Survey, the Dark Energy Spectroscopic Instrument, and the Vera C. Rubin Observatory, will enable many of these science goals, as well as further cross-wavelength analyses of large-scale structure in the Universe \citep{DES-Y6, DESI-DR2-1, DESI-DR2-2, Rubin-WhitePaper}. The observatory is expected to operate through the mid-2030s with a number of enhancements: the LAT has recently been upgraded to effectively double its mapping speed, while additional SATs will be deployed to improve sensitivity to the tensor-to-scalar ratio \citep{ASO-Abitbol_2025,MoreSATS-AbrilCabezas_2025}.

The science targets of the Simons Observatory require maps of the millimeter-wavelength sky with an unprecedented combination of area and sensitivity across a wide range of microwave frequencies \citep{SOScienceGoals-Ade_2019,ASO-Abitbol_2025}. To meet these needs, nearly 100,000 photon-noise-limited superconducting transition-edge sensor (TES) detectors have been deployed across the observatory's first four telescopes \citep{SATDeployment-Galitzki_2024,SATCommissioning_Harrington2026,LATRDark-Bhandarkar_2025}. These detectors cover a range of frequencies in order to provide sensitivity both to CMB signal and to galactic and extragalactic signals in the foreground. The LAT contains 63,000 optically-coupled TES detectors: 41,000 have mid-frequency (MF) passbands centered near 90\;GHz and 150\;GHz, 21,000 have ultra-high-frequency (UHF) passbands centered near 220\;GHz and 280\;GHz, and 700 have low-frequency (LF) passbands centered near 30\;GHz and 40\;GHz. The three SATs use the same readout and detector components as the LAT. Two of the three SATs each contain 12,000 optically-coupled TES detectors with MF passbands, while one of the SATs contains the same number of detectors with UHF passbands. The Simons Observatory requires a multiplexing scheme that can effectively read out this large number of cryogenic detectors. This is achieved using microwave superconducting quantum interference device (SQUID) multiplexing, which uses a cryogenic multiplexing circuit contained within purpose-built focal-plane modules and the SLAC National Accelerator Laboratory microresonator radio frequency (SMuRF) warm electronics \citep{UFM-McCarrick_2021,SMuRF-Yu_2023}. This system achieves a multiplexing factor of 860 for optically-coupled detectors, which is more than 12x that of telescopes which use other paradigms such as time-division or frequency-division multiplexing \citep{UMUX-Dober_2021}.

This paper details the implementation of this multiplexing scheme for the Simons Observatory LAT, which represents the largest deployment of TES microwave frequency multiplexing to a single telescope to date. We report on initial performance of the cryogenic readout components and SMuRF systems from the on-sky commissioning of the LAT's 63,000 detectors. Earlier LAT deployment and dark commissioning studies are reported in \cite{LATRDark-Bhandarkar_2025,LATRDark-Haridas_2024}; and \cite{LATRDark-Satterthwaite_2024}. This paper is structured as follows: Section \ref{sec:latr} summarizes key parts of the LAT's cryogenic receiver, Section \ref{sec:umux} describes the microwave frequency multiplexing system, and Section \ref{sec:latr-installation} describes the installation of this system to the receiver. Section \ref{sec:det-ops} discusses the operations that are performed in order to collect data using the LAT's readout system, and Section \ref{sec:characterization} presents results from a number of measurements which characterize the performance of this readout system, with comparisons to pre-deployment specifications, where appropriate.

\section{Large-Aperture Telescope Receiver} \label{sec:latr}

The LAT is a 6\;m reflective telescope with its primary and secondary mirrors in a crossed-Dragone configuration. Incident light is fed into a 2.3\;m-diameter LAT receiver (LATR) which achieves a diffraction-limited field of view of 7.8\degree at 150\;GHz \citep{Optics-Dicker_2018,LATOptics-Gudmundsson_21}. The focal plane, with its 63,000 sensors, is housed at 100\;mK, while ambient temperature is closer to 300\;K. The receiver therefore employs nested radiation shields at 300\;K, 80\;K, 40\;K, 4\;K, and 1\;K while en route to 100\;mK, which are cooled using five pulse tubes and a dilution refrigerator \citep{LATR-Xu_2020,LATRDesign-Zhu_2021}.

Light travels to the detectors through modular optics tubes (OTs), which contain a series of compartmentalized lenses and filters stepping from 4\;K to 100\;mK \citep{OTTesting-Harrington_2020,OTs-Sierra_2025}. At the 100\;mK stage, each OT has three universal focal-plane modules (UFMs), each measuring approximately 150\;mm across, which house the detectors and multiplexing chips \citep{UFM-McCarrick_2021}. Individual MF and UHF UFMs are designed to contain 1,720 optically-coupled detectors, while the LF UFMs, which have significantly larger feedhorns, are designed to contain 236 optically-coupled detectors each. Radio frequency (RF) and direct current (DC) cabling to facilitate readout of these detectors traverses the radiation shields and is brought to 300\;K via feedthrough bulkheads, at which point they connect to warm readout electronics \citep{URH-Moore_2022}.

This work presents measurements performed when the LAT contained its full suite of 13 OTs. These OTs were installed in stages with its first six installed by early 2024, and seven more installed as part of the telescope's upgrade in early 2026 \citep{ASO-Abitbol_2025}. Eight of these OTs contain MF detectors, four contain UHF detectors, and one contains LF detectors.

\section{Microwave Frequency Multiplexing} \label{sec:umux}

Across the Simons Observatory's current suite of telescopes, signals from the cryogenic detectors are measured using microwave frequency multiplexing $\left(\mu\text{mux}\right)$ \citep{SQUIDMUX-Irwin_2004,uMUX-Mates_2011}. This system consists of two key parts: a cryogenic circuit contained within the UFMs, and the SMuRF systems of warm electronics.

Each UFM contains a wafer of feedhorn-coupled orthomode transducers which provide sensitivity to the polarization of incoming radiation \citep{UMMAssembly-Healy_2020,UFM-McCarrick_2021,UHFUFM-Healy_2022,UFHFab-Duff_2024}. Pixels are dichroic; on-chip bandpass-defining filters determine the frequency dependence of the individual detectors to various frequency bands. Power incident to the UFMs is deposited onto DC-voltage-biased aluminum manganese TES detectors with a target critical temperature of 160\;mK and an operating temperature of around 100\;mK. The incident power induces a change in their resistances, which in turn alters the current through their circuits. These detector circuits are inductively coupled to RF SQUIDs on a series of multiplexing chips furnished to universal $\mu\text{mux}$ modules (UMMs), which are also packaged in the UFMs, so that the deposition of incident power causes a change in the flux through these SQUIDs \citep{UMUX-Dober_2021}. The SQUIDs are also coupled to a sawtooth flux ramp signal, which linearizes the otherwise periodic response of the SQUIDs. The SQUIDs screen inductors coupled to quarter-wave co-planar waveguide resonators on the UMMs with unique resonant frequencies in the 4\;GHz to 6\;GHz microwave range. The final effect of changing incident power on the TES detectors, then, is a perturbation of the phase of the flux-ramp-modulated frequencies of these resonators. A schematic of this circuit is shown in Figure \ref{fig:umux-schematic}.

\begin{figure}[t!]
\centering
\resizebox{\textwidth}{!}{
    \begin{tikzpicture}[
    scale=1.0,
    every node/.style={font=\small},
    /tikz/circuitikz/bipoles/thickness=1,
    tes/.style={rectangle, draw, thick, minimum width=8mm, minimum height=5mm},
    cylinder/.style={draw, thick, rounded corners=2pt, minimum width=6mm, minimum height=14mm, fill=gray!15}
    ]
     s
    \draw[thick] (-1.5,8.5) node[left] {RF in} 
        to[short, o-] (0,8.5) -- (18,8.5)
        to[short, -o] (19.5,8.5) node[right] {RF out};
    
    \draw (-1.5, 6.0) node[left] {Flux ramp}
        to[short, o-] (0,6.0) -- (1.4, 6.0)
        arc[start angle=180, end angle=0, radius=0.1cm]
        -- (2.6, 6.0);
    
    \draw (-1.5, 1.7) node[left] {TES bias}
        to[short, o-] (0,1.7) -- (2.6, 1.7);
    
    \begin{scope}[local bounding box=pixel1]
    
    \draw (1.5,8.5) to[C] (1.5,7.6) coordinate (cap1bot);
    \node[cylinder, anchor=north] (cyl1) at (cap1bot) {};
    \node[font=\footnotesize] at (cyl1.center) {$f_1$};
    \draw (cyl1.south) -- (1.5, 5.4);
    \draw (1.5, 5.4) to[L] (1.5, 4.2) coordinate (res1bot);
    \node[ground] at (res1bot) {};
    
    \draw (2.6, 6.0) to[L] (4.1, 6.0);
    
    \coordinate (sq1tl) at (2.6, 5.4);
    \coordinate (sq1tr) at (4.1, 5.4);
    \coordinate (sq1br) at (4.1, 4.2);
    \coordinate (sq1bl) at (2.6, 4.2);
    \draw[thick] (sq1tl) -- (sq1tr);
    \draw[thick] (sq1tr) to[barrier] (sq1br);
    \draw[thick] (sq1br) -- (sq1bl);
    \draw[thick] (sq1bl) -- (sq1tl);
    \node[font=\footnotesize, align=center] at (3.35, 4.8) {RF\\SQUID$_1$};
    
    \draw (2.6, 3.4) coordinate (Lin1L) 
        to[L] (4.1, 3.4) coordinate (Lin1R);
    
    \draw (Lin1L) -- (2.6, 2.9);
    \draw (2.6, 2.9) node[tes, anchor=north] (TES1) {TES$_1$};
    \draw (TES1.south) -- (2.6, 1.7);
    \draw (Lin1R) -- (4.1, 2.9);
    \draw (4.1, 2.9) -- (4.1, 1.7);
    
    \draw (2.6, 1.7) to[R, l=$R_\text{sh}$, /tikz/circuitikz/bipoles/length=10mm] (4.1, 1.7);
    
    \end{scope}
    
    \begin{scope}[xshift=5cm, local bounding box=pixel2]
    
    \draw (1.5,8.5) to[C] (1.5,7.6) coordinate (cap2bot);
    \node[cylinder, anchor=north] (cyl2) at (cap2bot) {};
    \node[font=\footnotesize] at (cyl2.center) {$f_2$};
    \draw (cyl2.south) -- (1.5, 5.4);
    \draw (1.5, 5.4) to[L] (1.5, 4.2) coordinate (res2bot);
    \node[ground] at (res2bot) {};
    
    \draw (2.6, 6.0) to[L] (4.1, 6.0);
    
    \coordinate (sq2tl) at (2.6, 5.4);
    \coordinate (sq2tr) at (4.1, 5.4);
    \coordinate (sq2br) at (4.1, 4.2);
    \coordinate (sq2bl) at (2.6, 4.2);
    \draw[thick] (sq2tl) -- (sq2tr);
    \draw[thick] (sq2tr) to[barrier] (sq2br);
    \draw[thick] (sq2br) -- (sq2bl);
    \draw[thick] (sq2bl) -- (sq2tl);
    \node[font=\footnotesize, align=center] at (3.35, 4.8) {RF\\SQUID$_2$};
    
    \draw (2.6, 3.4) coordinate (Lin2L) 
        to[L] (4.1, 3.4) coordinate (Lin2R);
    
    \draw (Lin2L) -- (2.6, 2.9);
    \draw (2.6, 2.9) node[tes, anchor=north] (TES2) {TES$_2$};
    \draw (TES2.south) -- (2.6, 1.7);
    \draw (Lin2R) -- (4.1, 2.9);
    \draw (4.1, 2.9) -- (4.1, 1.7);
    
    \draw (2.6, 1.7) to[R, l=$R_\text{sh}$, /tikz/circuitikz/bipoles/length=10mm] (4.1, 1.7);
    
    \end{scope}
    
    \begin{scope}[xshift=12cm, local bounding box=pixelN]
    
    \draw (1.5,8.5) to[C] (1.5,7.6) coordinate (capNbot);
    \node[cylinder, anchor=north] (cylN) at (capNbot) {};
    \node[font=\footnotesize] at (cylN.center) {$f_N$};
    \draw (cylN.south) -- (1.5, 5.4);
    \draw (1.5, 5.4) to[L] (1.5, 4.2) coordinate (resNbot);
    \node[ground] at (resNbot) {};
    
    \draw (2.6, 6.0) to[L] (4.1, 6.0);
    
    \coordinate (sqNtl) at (2.6, 5.4);
    \coordinate (sqNtr) at (4.1, 5.4);
    \coordinate (sqNbr) at (4.1, 4.2);
    \coordinate (sqNbl) at (2.6, 4.2);
    \draw[thick] (sqNtl) -- (sqNtr);
    \draw[thick] (sqNtr) to[barrier] (sqNbr);
    \draw[thick] (sqNbr) -- (sqNbl);
    \draw[thick] (sqNbl) -- (sqNtl);
    \node[font=\footnotesize, align=center] at (3.35, 4.8) {RF\\SQUID$_N$};
    
    \draw (2.6, 3.4) coordinate (LinNL) 
        to[L] (4.1, 3.4) coordinate (LinNR);
    
    \draw (LinNL) -- (2.6, 2.9);
    \draw (2.6, 2.9) node[tes, anchor=north] (TESN) {TES$_N$};
    \draw (TESN.south) -- (2.6, 1.7);
    \draw (LinNR) -- (4.1, 2.9);
    \draw (4.1, 2.9) -- (4.1, 1.7);
    
    \draw (2.6, 1.7) to[R, l=$R_\text{sh}$, /tikz/circuitikz/bipoles/length=10mm] (4.1, 1.7);
    
    \end{scope}
    
    \draw (4.1, 6.0) -- (6.4, 6.0)
        arc[start angle=180, end angle=0, radius=0.1cm]
        -- (7.6, 6.0);
    \draw (9.1, 6.0) -- (13.4, 6.0)
        arc[start angle=180, end angle=0, radius=0.1cm]
        -- (14.6, 6.0);
    \draw (16.1, 6.0) -- (18, 6.0)
        to[short, -o] (19.5, 6.0);
    
    \draw (4.1, 1.7) -- (7.6, 1.7);
    \draw (9.1, 1.7) -- (14.6, 1.7);
    \draw (16.1, 1.7) -- (18, 1.7)
        to[short, -o] (19.5, 1.7);
    
    \node at (10.5, 6.9) {\Large $\cdots$};   
    \node at (10.5, 4.8) {\Large $\cdots$};   
    \node at (10.5, 2.65) {\Large $\cdots$};  
    
    \end{tikzpicture}
}
\caption{Schematic of the cryogenic microwave frequency multiplexing circuit, showing a selection of TES detectors which share a common DC TES bias line, as well as their SQUIDs which share a single flux ramp. Microwave resonator frequencies are labeled with $f_i$ and shunt resistances are labeled with $R_\text{sh}$. TES detectors deployed to the Simons Observatory have a target normal resistance of 8\;m$\Omega$ and shunt resistors have a target resistance of 400\;$\mu\Omega$. TES bias lines have an effective resistance of approximately 16\;k$\Omega$ due to inline room-temperature variable resistors. Open SQUID and dark channels are not shown.
\label{fig:umux-schematic}}
\end{figure}

Each MF and UHF UFM is designed to contain 1,848 microwave resonators, 1,820 of which are coupled to SQUIDs. Of these SQUIDs, 1,756 are coupled to TES circuits: 1,720 optically-coupled channels and 36 dark channels used for calibration. This set of 1,756 detectors is split into two halves with a dedicated flux ramp and six DC TES bias lines on each half; a single transmission line reads out each half of the UFM, therefore the multiplexing factor for optically-coupled TES detectors is 860. The LF UFMs each contain 264 resonators, 260 of which are coupled to SQUIDs, and 244 of which are coupled to detectors, with 236 optically coupled. These detectors are split across four DC TES bias lines and a single flux ramp.

\begin{sidewaysfigure}
    \centering
    \includegraphics[width=\textheight]{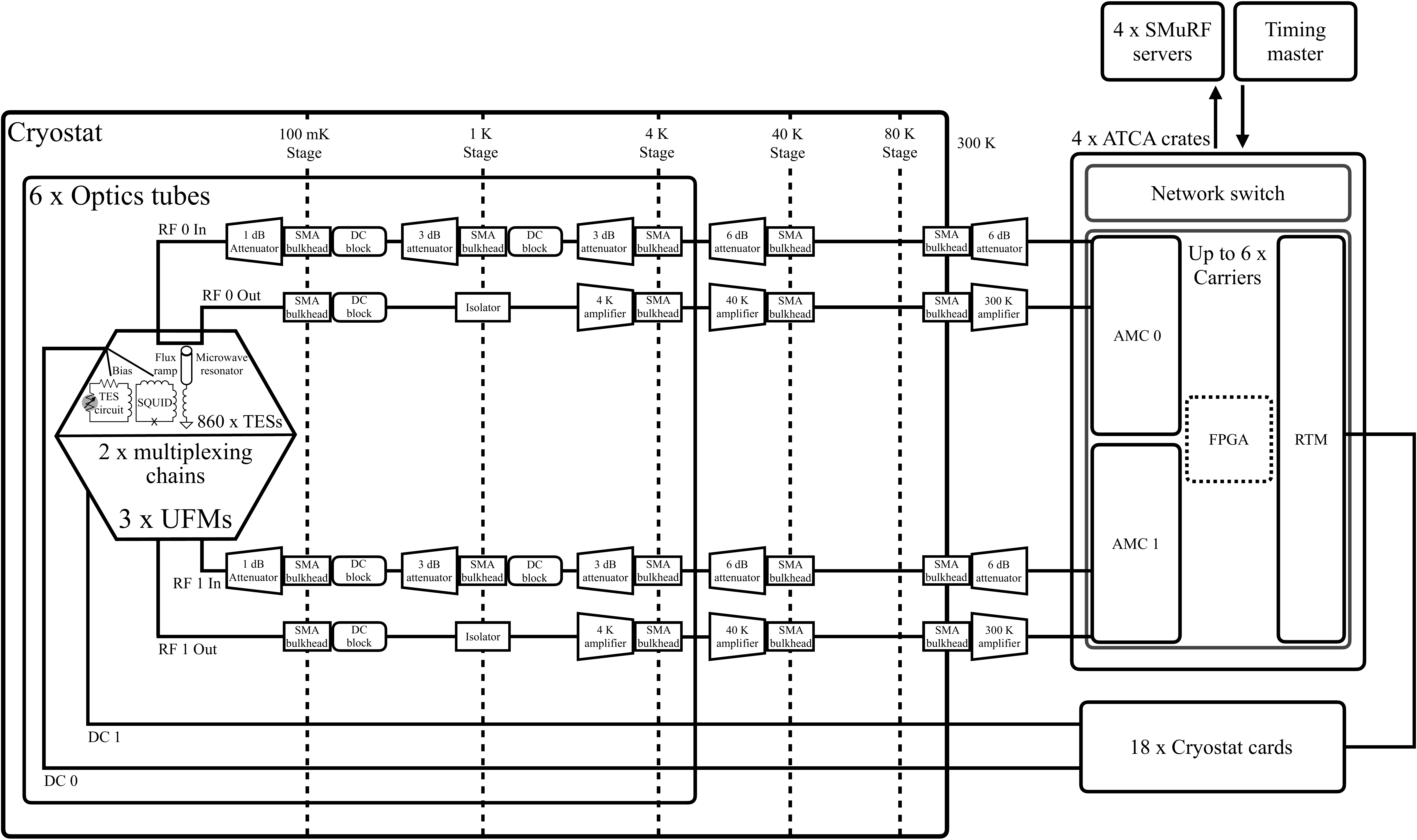}
    \caption{Block diagram showing the RF and DC chains of the LATR, from the TES detectors to the data acquisition servers. The figure shows the most relevant components, including the RF attenuators and amplifiers. The figure omits DC wiring for cold amplifier biasing. One representative TES circuit is shown inside one MF or UHF UFM (of a total of 39 modules across all passbands) with its two RF and DC chains. Note that each UFM's DC lines share a single cable from the cryostat cards that gets broken out at various stages. Not drawn to scale.}
    \label{fig:network-diagram}
\end{sidewaysfigure}

The full RF and DC chains for the LATR, from the TES detectors to the data acquisition servers, are shown in Figure \ref{fig:network-diagram}. Signals from the warm readout electronics enter the cryostat via universal readout harnesses (URHs) which provide the interface between the ambient 300\;K outside the cryostat and the 4\;K stage which houses the OTs \citep{RFChain-Rao_2020,URH-Moore_2022}. A single URH carries signals for up to 24 individual RF and DC chains, so four URHs are used to read out the 13 OTs discussed in this work. RF excitation tones for the microwave resonators pass from the SMuRF electronics to the URH through a warm coaxial cable and a 6\;dB attenuator at 300\;K, before entering the cryostat directly into the 40\;K stage via a hermetic SMA bulkhead adapter. Inside the cryostat, a cupronickel coaxial cable carries the tones to 4\;K, with a second 6\;dB attenuator and a second SMA bulkhead adapter at the 40\;K stage, and a final 3\;dB attenuator at the 4\;K stage. Here, an isothermal copper coaxial cable carries the tones from the URH into an OT en route to the resonators at 100\;mK. After interrogating the resonators and exiting the UFM, the power deficit signal from these tones ($S_{21}$) passes back to the 4\;K stage through a superconducting niobium-titanium coaxial cable and an isolator\footnote{Low Noise Factory part number ISC4\_8A.} at 1\;K. The signals then travel through a low-noise amplifier (LNA) at 4\;K\footnote{Low Noise Factory part number LNC4\_8C.}, an SMA bulkhead to 40\;K, a second LNA at 40\;K\footnote{Custom-built amplifier; see \cite{URH-Moore_2022}.}, a hermetic SMA bulkhead to 300\;K, a third LNA at 300\;K\footnote{Mini-Circuits part number ZX60-83LN12+.}, and finally a warm coaxial cable which enters back into the SMuRF system. DC TES bias and alternating-current flux ramp signals pass from the SMuRF through a 100-pin Small Computer System Interface (SCSI) cable to a cryostat card. From the cryostat card, the signals pass through a 50-pin subminiature-D cable before entering the URH where they are broken into a 51-pin micro-D cable at the 40\;K and 4\;K stages, then transitioned into a 37-pin micro-D cable before entering the UFM.

The SMuRF electronics provide the warm system for reading out data from the UFMs, carried through these URHs \citep{EarlySMuRF-Henderson_2018,SMuRF-Yu_2023}. Each SMuRF system consists of a field-programmable gate array (FPGA)-driven carrier board, two Advanced Mezzanine Cards (AMCs), a Rear Transition Module (RTM), and a cryostat card. The carrier board interfaces with the system's other components: the AMCs generate and demodulate RF tones in a 4-6\;GHz range, and the RTM generates the DC signals for the flux ramp, the cryogenic amplifier biases, and the TES detector biases that enter the cryostat by way of the cryostat card. The carrier boards and their AMCs and RTMs are installed in Comtel C07 Advanced Telecommunications Computing Architecture (ATCA) crates alongside a Vadatech ATC807 network switch which interfaces with a dedicated data acquisition server via Ethernet and optical fiber links, and a timing system which synchronizes with an observatory-wide 122.88\;MHz clock. Cryostat cards are housed inside separate aluminum enclosures which provide further isolation from sources of noise, and these enclosures are bolted directly to the exterior of the LATR. Up to six SMuRF systems and one network switch are installed in each ATCA crate. Each SMuRF system drives one MF or UHF UFM with each AMC producing RF tones for a single UFM half and each cryocard splitting DC signals between the two halves; one fully-loaded seven-slot ATCA crate can therefore read out six such UFMs, or two OTs. As LF UFMs have fewer detectors, a single carrier, AMC, RTM, and cryostat card system can read out three such UFMs, with their RF chains connected in serial and their cryostat card splitting the DC signals across the three modules.

As incident power changes the resistance of an individual TES detector, the SMuRF system is designed to detect the transduced change in the associated microwave resonator's characteristics via RF probe tones generated by an AMC. The resonant frequencies of these resonators are oscillating due to the linearizing flux ramp signal generated by the RTM, so this change in incident signal induces a change in phase of the resonant frequency modulation. This shift in phase is demodulated at the flux ramp frequency, and it is recorded by data acquisition servers for offline analysis.

\section{Installation to the LATR} \label{sec:latr-installation}

\begin{figure*}[t!]
\centering
\includegraphics[width=0.8\textwidth]{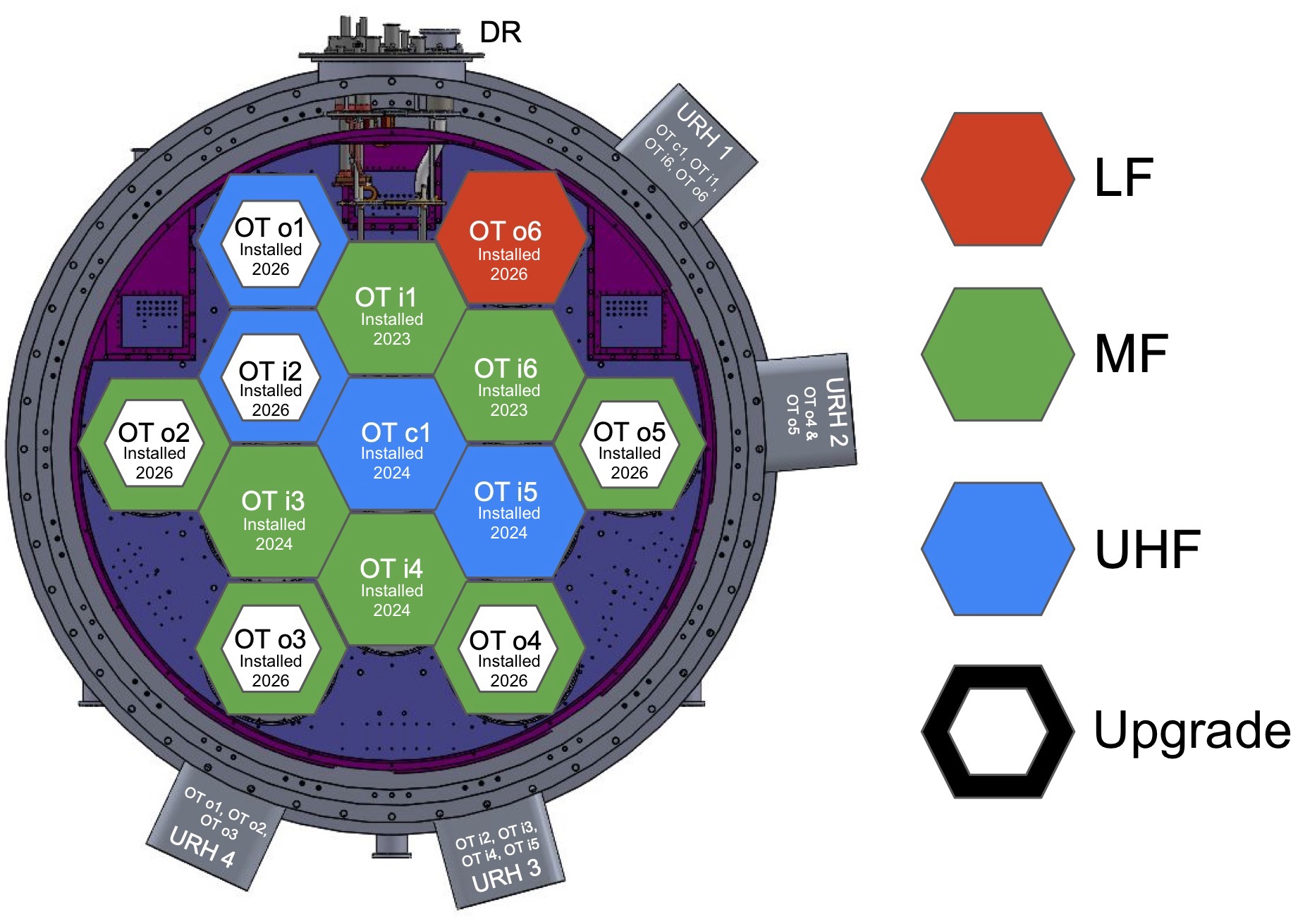}
\caption{Rear-view schematic of the LATR with the positions of its OTs, URHs, and dilution refrigerator (DR) shown. Passbands and installation dates of the OTs are marked. ``Upgrade" refers to the six OTs produced as part of the LAT's enhancement \citep{ASO-Abitbol_2025}. OTs are numbered in a counterclockwise fashion (when viewed from the rear) beginning from the top of the receiver, with the penultimate character of an OT identifier signifying whether it is part of the central (c), inner (i), or outer (o) ring.}
\label{fig:ot-pattern}
\end{figure*}

The LAT's OTs were deployed in phases. The instrument's first two OTs were installed in mid 2023, four more were installed in early 2024, and its final seven were installed in early 2026, leading to a total of 39 UFMs. UFMs were fabricated and characterized in North America, then in Chile they were installed on the gold-plated copper focal plane plates that house each OT's 100\;mK readout components and wiring to 4\;K, before the OTs were installed in the LATR \citep{UFMTesting-Wang_2022,MFUFM-Dutcher_2023,UHFUFM-Dutcher_2025}. OTs are arranged in concentric rings within the LATR as shown in Figure \ref{fig:ot-pattern}, which also indicates the passband of each OT.

\begin{figure*}[t!]
\centering
\includegraphics[width=0.8\textwidth]{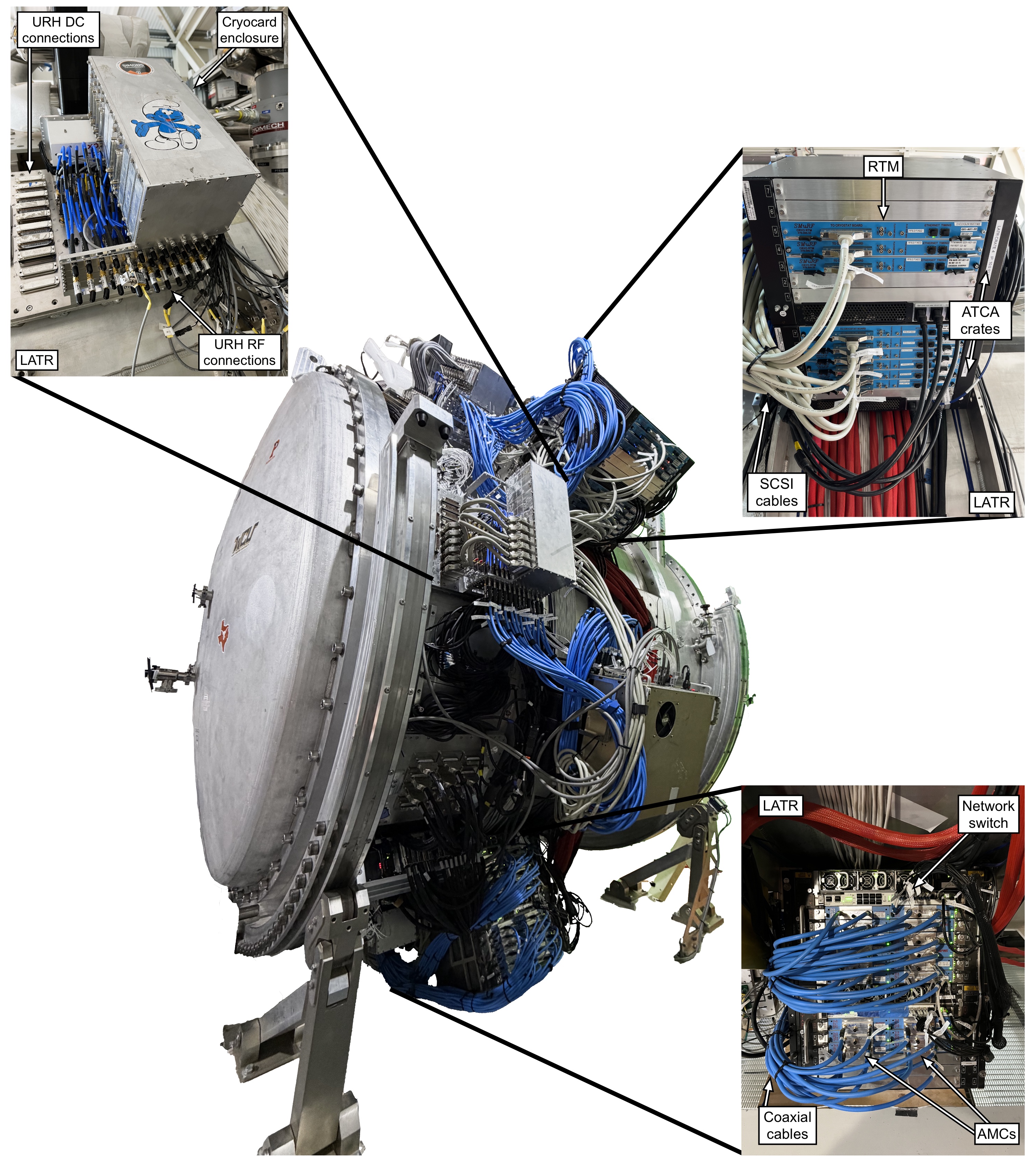}
\caption{Cutout of the LATR as viewed from the rear, with insets showing various SMuRF components. Top left: a cryostat card enclosure and its URH connections, without cables installed. Top right: the back of two ATCA crates of SMuRF systems, showing their RTMs and SCSI cables. Bottom right: the front of two ATCA crates of SMuRF systems, showing their network switches, AMCs, and coaxial cables.}
\label{fig:smurf-latr}
\end{figure*}

Alongside these UFMs, a commensurate number of SMuRF systems have also been installed in the LAT at the same cadence in order to enable the warm readout of these detector modules. The seven ATCA crates which host these SMuRF systems are mounted directly to the LATR. Up to two ATCA crates are mounted adjacent to each URH, alongside a single aluminum enclosure which contains up to twelve cryostat cards for the neighboring pair of crates. Figure \ref{fig:smurf-latr} shows a cutout of the LATR, with insets showing a pair of ATCA crates and one cryostat card enclosure. The seven data acquisition servers that communicate with these SMuRF systems are mounted to a server rack two floors below; optical fiber links for fast data transmission are routed through a cable wrap to avoid tangling during boresight rotation.

During the deployment and commissioning period presented in this paper, we addressed a number of hardware-related issues with the readout chain. The receiver cabin which houses the LATR is kept isothermal with the ambient environment on Cerro Toco so that the LATR structure experiences the same diurnal temperature changes as the mirror structure and optical stability is maintained \citep{CCAT-Parshley_2018}. While previous work has established that these diurnal temperature changes do not introduce a systematic signal contamination via changes to the physical lengths and dielectric permittivities of the SMuRF systems' warm coaxial cables, the weather of the Atacama Desert has posed a challenge nonetheless \citep{ToneTracking-SilvaFeaver_2022,PhaseDrift-Satterthwaite_2025}. Many of the commercial components for the SMuRF systems, such as Small Form-Factor Pluggable (SFP) fiber modules and ATCA crate power supply units, are not designed for operation below 0\degree\;C, though the receiver cabin spends a significant amount of time below this temperature. While we have been able to mitigate temperature-related failures of some of these components by procuring ``industrial" versions, which are rated to significantly lower temperatures, we have had to actively monitor for failures of other components and replace as necessary from an on-site cache of spares. We have also employed engineering measures to protect the SMuRF components from precipitation on Cerro Toco amid the receiver cabin's exposure; weather sealing around the structure surrounding the cabin has prevented moisture-related failures of SMuRF components. Finally, we also experienced installation issues within the cryostat, such as swapped coaxial cables and a small number of failed cryogenic amplifiers. These issues were addressed by warming and opening the LATR in situ, and moving or replacing the relevant components.

We have improved the strain relieving of SMuRF cables to avoid disengagement caused by LATR rotation and LAT motion. Strain at the connection point under the weight of the warm cables can cause disengagement, particularly of the DC SCSI cables which have a notably thin interface and a short leg length. To prevent this, we torque the connectors' screwlocks evenly into their mating jackposts to ensure proper, durable engagement. Similarly, we have implemented a screwlock system for the DC subminiature-D cables which engages directly with the cryostat's exterior shell. Solving these engagement issues and sourcing components which are robust to the ambient environment are key factors which have enabled us to sustain operation of the readout system with minimal physical intervention.

\section{Readout Operations}
\label{sec:det-ops}

We perform a series of readout setup operations to prepare for data acquisition from the UFMs when they are at their base temperature of 100\;mK, which we detail below. These operations are implemented by the public \texttt{pysmurf}\footnote{\href{https://github.com/slaclab/pysmurf}{https://github.com/slaclab/pysmurf}
} and \texttt{sodetlib}\footnote{\href{https://github.com/simonsobs/sodetlib}{https://github.com/simonsobs/sodetlib}
} software repositories.

\subsection{Resonator tuning}
\label{subsec:tuning}

\begin{figure*}[t!]
\centering
\includegraphics[width=\textwidth]{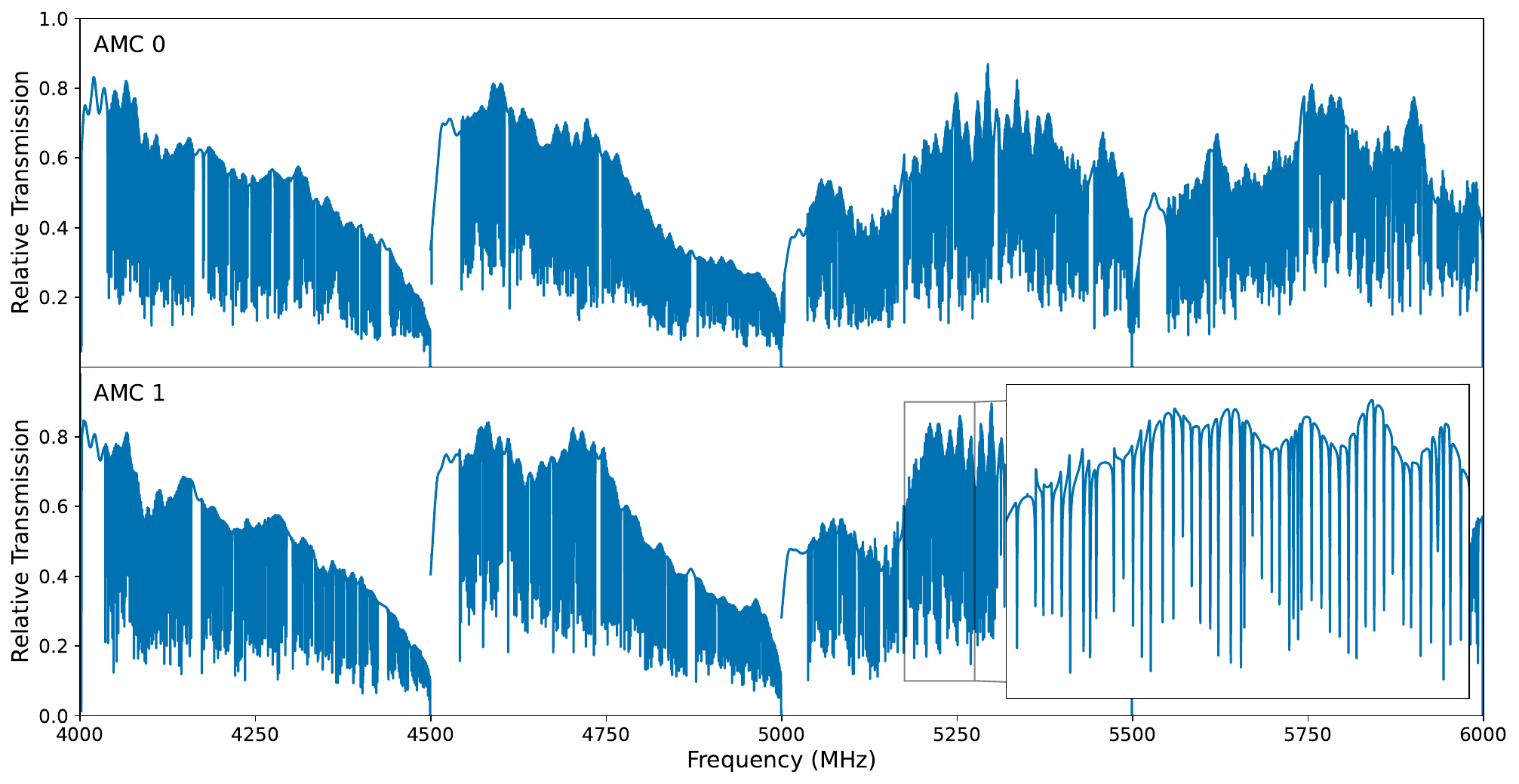}
\caption{Frequency response of the microwave resonators on a single UFM deployed to the LATR, as probed by the two AMCs of a single SMuRF system. Tones are generated in 500\;MHz-wide bands, leading to the jumps at 4,500\;MHz, 5,000\;MHz, and 5,500\;MHz. Inset shows the zoomed-in response over a 100\;MHz region.
\label{fig:tuneplot}}
\end{figure*}

\begin{figure*}[t!]
\centering
\includegraphics[width=\textwidth]{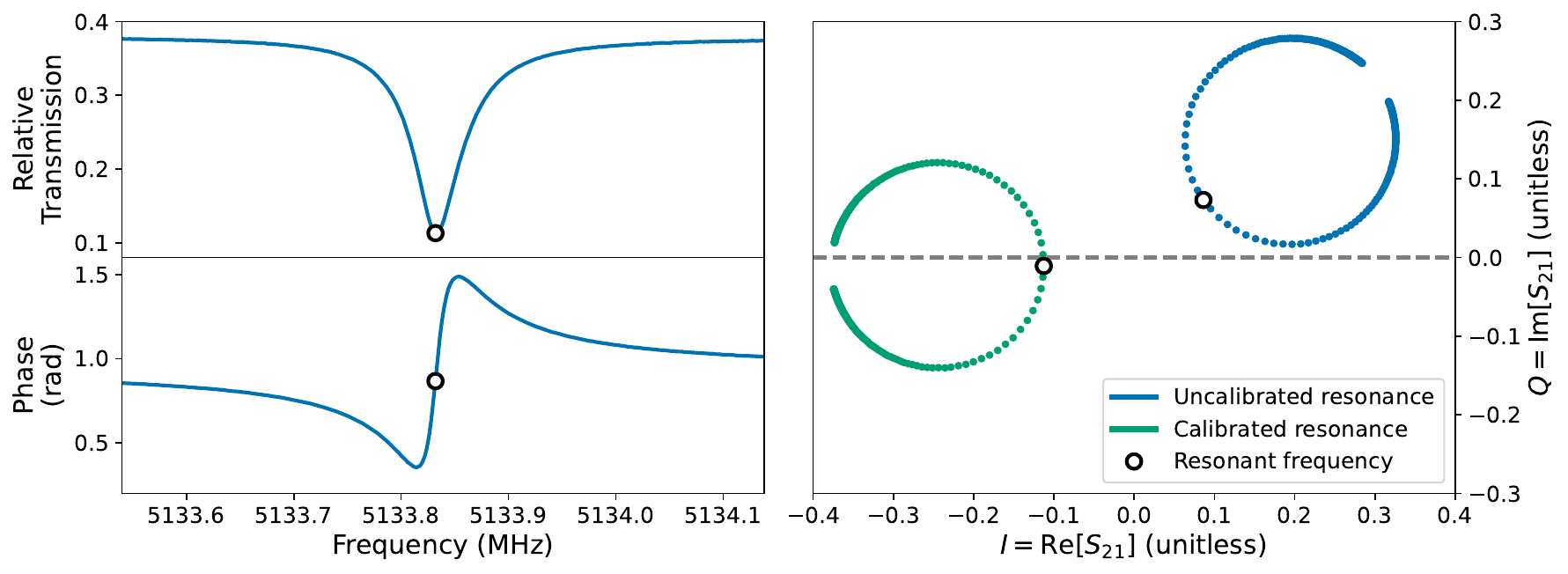}
\caption{Sample microwave resonator on a UFM deployed to the LATR. Left: amplitude and phase of the post-calibration resonance. Right: resonance shown in the $IQ$-plane before (blue) and after (green) calibration. Resonant frequency is marked with black circles.
\label{fig:resonance}}
\end{figure*}

We first bias the warm and cryogenic amplifiers to their respective drain currents, and identify the frequency location of each of the microwave resonators. The latter process is referred to as ``tuning," and it first sweeps tones across the 4-6\;GHz range probed by each AMC. Individual tones are generated within 500\;MHz-wide bands by dedicated digital-to-analog converters (DACs) and upconverted to the desired frequency via mixing with local oscillators (LOs). As these tones are swept across the SMuRF's bands, the microwave resonators become apparent as characteristic dips in the response $\left(S_{21}\right)$, which is demodulated by SMuRF after being downmixed with the LOs and processed by analog-to-digital converters (ADCs). Figure \ref{fig:tuneplot} shows the magnitude of this response on a single UFM installed in the LATR as probed by two AMCs, with a selection of the resonators shown in the zoomed-in inset.

Individual resonances are then identified by searching for local minima in the magnitude of the response. These minima provide estimates of the frequency locations of the resonators; a finer frequency sweep and finally a gradient descent around the minima refines these estimates. Using this finer frequency sweep, software identifies the locations of the resonant frequencies. Transmission power is then converted into in-phase $\left(I=\text{Re}\left[S_{21}\right]\right)$ and quadrature $\left(Q=\text{Im}\left[S_{21}\right]\right)$ components in order to compute a calibrating rotation $\eta$ in the $IQ$-plane which enforces that changes to this frequency are purely imaginary. Small perturbations about a resonant frequency therefore produce changes in probe tone response which are, to first order, entirely in $Q$. The changes are then converted from voltages measured by the ADCs to frequency shifts using this calibration, which is shown in Figure \ref{fig:resonance}.

This initial resonator tuning process is run in parallel across SMuRF systems and it typically takes 30-45\;minutes, with the fine frequency sweep and gradient descent about resonator minima driving the execution time. As resonator locations, in the absence of a flux ramp or incident TES signal, do not appreciably shift during operation, we run this tuning step once per cool-down of the LATR. As warm-ups of the LATR are rare, the execution time of this operation does not noticeably impact the LAT's mapping speed.

\subsection{Resonator tracking}
\label{subsec:tracking}

\begin{figure*}[t!]
\centering
\includegraphics[width=\textwidth]{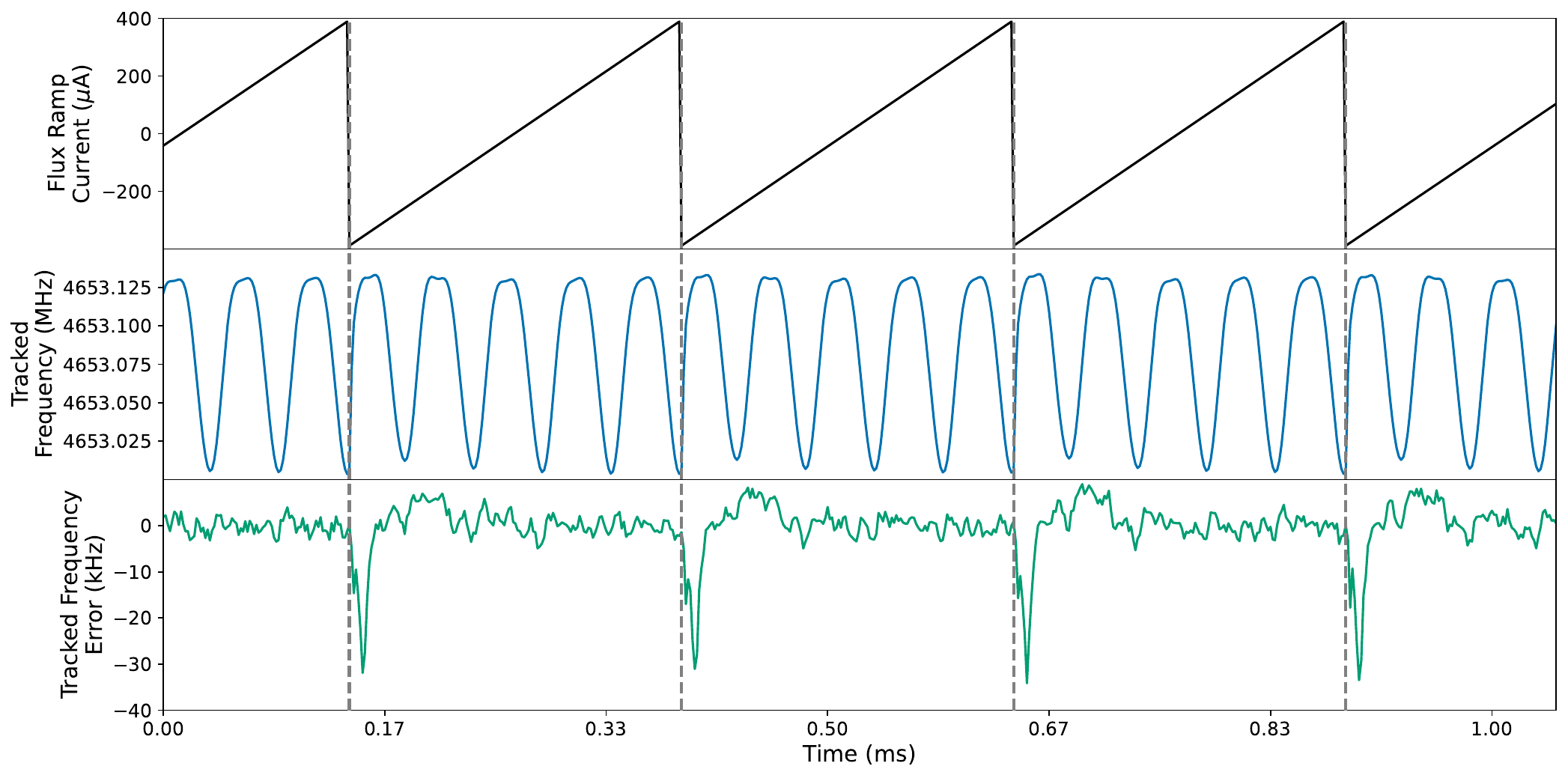}
\caption{Frequency response of a single microwave resonator on a UFM deployed to the LATR as it is flux ramped. Vertical dashed lines indicate the times of flux ramp resets. Top: flux ramp current, scaled using design parameters for Simons Observatory SQUIDs. Middle: tracked frequency response of the resonator. Bottom: error on the tracked frequency response of the resonator.
\label{fig:trackplot}}
\end{figure*}

The flux ramp that linearizes the SQUIDs' responses drives a periodic oscillation of the resonators' frequencies at a specified rate. To maximize sensitivity, the SMuRF system continuously updates the probe tone interrogating each resonator to remain near the resonant frequency. This process is enabled by tone-tracking firmware implemented on the SMuRF FPGAs. This begins by generating the system's flux ramp, whose frequency must be faster than that of the TES signal; for the Simons Observatory's detectors with millisecond time constants, the RTMs generate a flux ramp at $4\;\text{kHz}$, as shown in the top panel of Figure \ref{fig:trackplot}. We target a flux ramp amplitude equivalent to five flux quanta in the SQUIDs as a flux ramp modulation frequency of 20\;kHz has been shown empirically to minimize RF phase noise \citep{Thesis-SilvaFeaver_2023}. This creates the five oscillations of microwave resonator frequency per flux ramp shown in the middle panel of Figure \ref{fig:trackplot}.

Firmware implemented on the SMuRF then uses an algorithm akin to stochastic gradient descent to fit the sum of three sine and cosine harmonics to the frequency response of each resonator (see \cite{SMuRF-Yu_2023}). The residual error from this model, which is minimized by the firmware, is shown in the bottom panel of Figure \ref{fig:trackplot}. Note that the firmware omits regions close to the flux ramp resets from the fit due to the transient response of the flux ramp circuit; including these regions could otherwise lead to a systematic bias. During this process, the probe tones produced by the AMCs are updated rapidly to ensure that they follow the oscillating resonant frequencies. As the TES detector signal changes, this induces a phase change in the tracked oscillation, which is measured from the fit over the length of a single flux ramp period.

To initialize the resonator tracking, data is recorded at 600\;kHz with the flux ramp turned on and this feedback process enabled in order to determine the optimal tracking parameters. This happens in three stages. The first execution determines the common flux ramp amplitude necessary to induce five flux quanta in the SQUIDs. The tones which probe resonators with anomalous peak-to-peak frequency changes upon this flux ramping (less than 10\;kHz or greater than 200\;kHz) or anomalously high peak-to-peak errors on the tracked frequency swings (greater than 50\;kHz) are then disabled. The second execution refines the flux ramp amplitude without being biased by poorly-performing SQUIDs or low-quality microwave resonators. Finally, the tones probing these anomalous channels are re-enabled and the tracking algorithm is run a final time with fixed flux ramp parameters from the second execution in order to yield a final fit to each resonator.

This tracking initialization process typically takes about 5\;minutes and it is run after every tune of the system, which is roughly once per cool-down of the LATR.

\subsection{$I{-}V$ curves}
\label{subsec:iv}

\begin{figure*}[t!]
\centering

\begin{subfigure}[b]{0.335\textwidth}
    \centering
    \includegraphics[width=\textwidth]{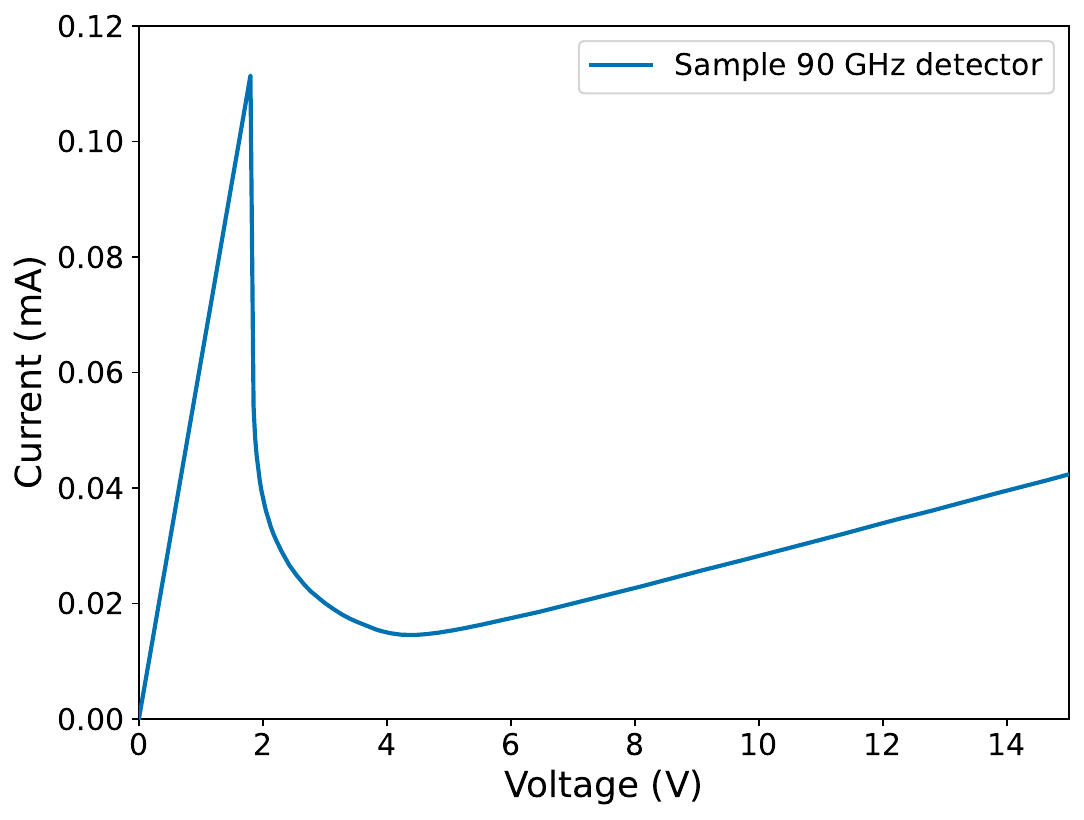}
    \caption{}
    \label{fig:iv-raw}
\end{subfigure}
\begin{subfigure}[b]{0.625\textwidth}
    \centering
    \includegraphics[width=\textwidth]{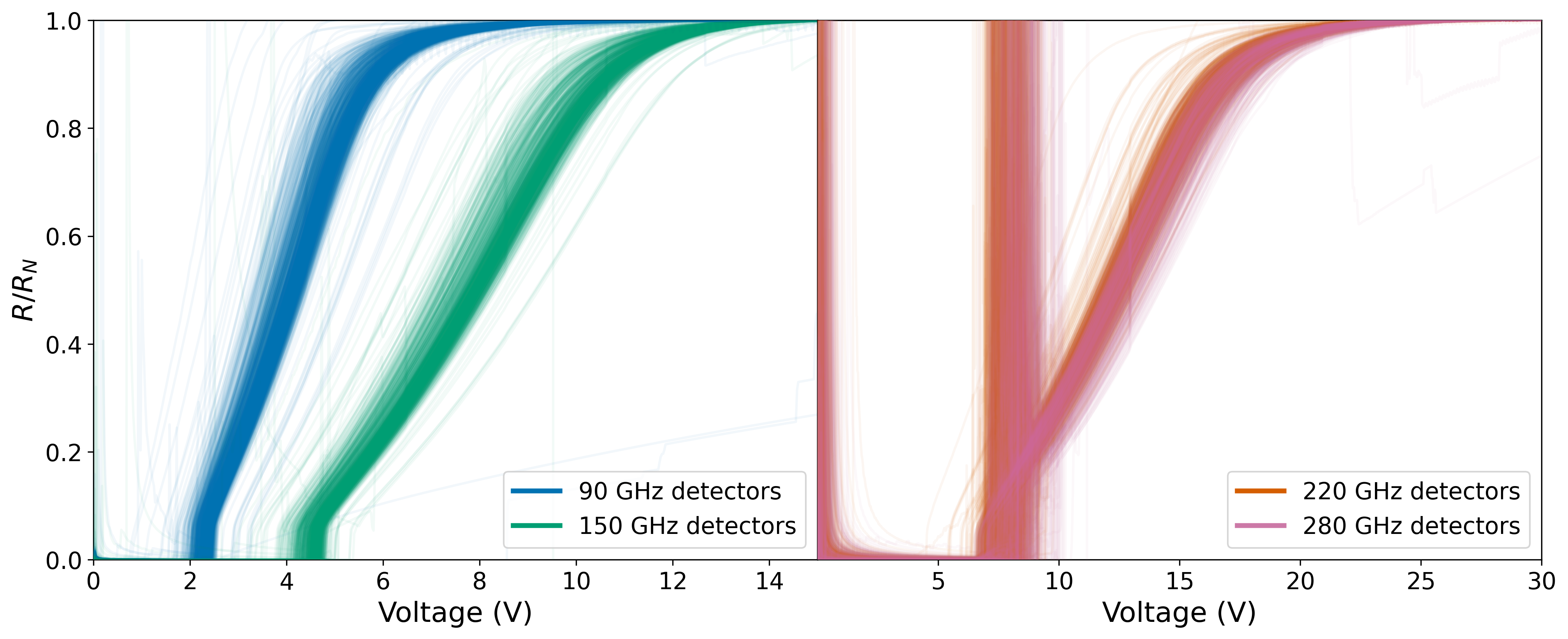}
    \caption{}
    \label{fig:ivs-scaled}
\end{subfigure}
\caption{Sample $I{-}V$ curves from two UFMs deployed to the LATR. (a) Response from a single 90\;GHz detector. (b) Responses from all detectors on each UFM with the vertical $\left(I\right)$ axis converted to $R$ and scaled by $R_N$. Curves from both MF and UHF UFMs are shown in the right-hand side plot. Note that the UHF detectors exhibit instability at low bias points, where oscillations of the current across the TES are stopped when low current drops the detector into its superconducting branch. This does not affect our measurements as we seek to bias these detectors to higher values of $R/R_N$.
\label{fig:ivs}}
\end{figure*}

Once the microwave resonators have been located and are being tracked, the next step is to DC-voltage bias their detectors into their superconducting transition states. A UFM's detectors are spread across a number of bias lines, with half of the bias lines dedicated to each passband. The biasing process first empirically maps detectors and resonators to their bias groups. The RTM sends a small voltage down the bias lines, then serially steps each bias line's voltage in a square wave. As the detectors remain superconducting at this low bias, the waveform changes the current through each TES detector circuit, therefore changing the flux through the SQUIDs, and inducing a response from the microwave resonators. Using this response, server-side data analysis then determines which resonators belong to each bias line.

Each detector's response to bias voltage is then characterized via $I{-}V$ curves. These curves are produced by applying a voltage bias across the detectors which exceeds their critical currents, therefore driving them from their superconducting states to a resistive regime. This voltage is then reduced so that the TES detectors are kept normal by their self-heating, before slowly stepping the voltage down so the detectors pass through their transitions and return to superconductivity. For LF and MF detectors, this is achieved by commanding an over-bias of 18\;V from the SMuRF, while for UHF detectors, an over-bias voltage of 35\;V is commanded\footnote{These biases are achieved by running SMuRF in ``high-current mode." Biases presented in this paper are scaled by the high-to-low-current ratio of about 5.7. For further discussion, see \cite{SMuRF-Yu_2023}.}. Given shunt resistances of 400\;$\mu\Omega$ and bias line resistances of approximately 16\;k$\Omega$, this leads to roughly 0.4\;$\mu$V and 0.8\;$\mu$V across the LF/MF and UHF detectors, respectively. In both cases, the RTM then steps the bias voltage down by 0.025\;V every 0.1\;s while measuring the response of each detector via its SQUID-coupled microwave resonator. This traces curves as shown in Figure \ref{fig:ivs}, where Figure \ref{fig:iv-raw} shows the current response of a single detector, and Figure \ref{fig:ivs-scaled} shows responses which are converted to TES resistance $R$, scaled to $R_\text{frac}$ using normal resistance $R_N$: $R_\text{frac}\equiv R/R_N$. Note that we observe instability in the UHF detectors around low bias points due to their fast time constants. However, this does not appreciably affect our measurement of the transition, and we seek to bias detectors closer to $R_\text{frac}\approx0.5$. This choice of $R_\text{frac}$ allows us to effectively balance stability of the bias point with effective electrothermal feedback, or loop gain, while also maintaining a margin to saturation. We bias detectors approximately to this point by selecting the median voltage which corresponds to $R_\text{frac}=0.5$ for all detectors on each bias line.

Changes in the precipitable water vapor (PWV) on Cerro Toco and changes in observing elevation lead to changes in optical detector loading, which alter the optimal TES bias voltages.  We therefore perform this calibration step and re-bias detectors to their optimal bias points often during observations. We collect $I{-}V$ curves, which take about 10 minutes, with every change in elevation, and at least every four hours in the absence of changes in elevation in order to capture changes in PWV. We find this to be an effective balance between maintaining optimal detector bias points and observing efficiency.

With detectors tuned, tracked, and biased, we stream data from the LATR. In order to achieve a sampling rate of 200\;Hz, which is motivated by the instrument's science goals, we down-sample data by 20x from the 4\;kHz flux ramp rate using a low-pass infinite impulse response filter. Future publications will discuss the measurement of detector parameters such as power-to-current responsivities and time constants.

\section{Readout Characterization} \label{sec:characterization}

During the commissioning phase of the LAT, we completed a number of measurements to characterize the performance of this readout system, which we present in this section. Results from on-sky optical characterization will be presented in a future publication.

\subsection{Resonator, SQUID, and detector count}
\label{subsec:mux-count}

The procedure detailed in Section \ref{sec:det-ops} provide measurements of the number of operational resonators, SQUIDs, and TES detectors across the LATR. We define these measurements to be the number of channels whose:

\begin{itemize}
    \item resonators are located as well-defined dips in the $S_{21}$ response of their corresponding UFM during tuning, as shown in the inset in Figure \ref{fig:tuneplot};
    \item resonators are tracked and exhibit a peak-to-peak change in resonant frequency between 10\;kHz and 200\;kHz;
    \item detectors are biased by producing an $I{-}V$ curve which passes through $R_\text{frac}=0.9$ and which shows a normal resistance $R_n$ between 2\;m$\Omega$ and 20\;m$\Omega$ (to filter anomalous non-physical values).
\end{itemize}

We present measurements of these quantities as a validation of the successful operability of the readout system as configured and deployed, and an assessment of its performance. We analyze observations during 1 May 2026 through 1 July 2026, as PWV values are generally low in this period of austral autumn and winter \citep{PWV-Cortes_2020}. For each UFM, we select the CMB observation with the highest count of biased optically-coupled detectors, where we define biased as $0.2<R_\text{frac}<0.8$, and we identify the detectors which are optically coupled using the algorithm described in \cite{DetMatch-Lashner_2024}. We then use the tuning, tracking, and $I{-}V$ curve acquisitions directly preceding these observations to measure the number of resonators, SQUIDs, and TES bolometers, respectively, as described above. Note that this is not a count of simultaneously biased bolometers during operation, as presented in \cite{SATCommissioning_Harrington2026}; such a measurement for the LAT will be presented in a future work.

\begin{figure*}[t!]
\centering
\includegraphics[width=0.8\textwidth]{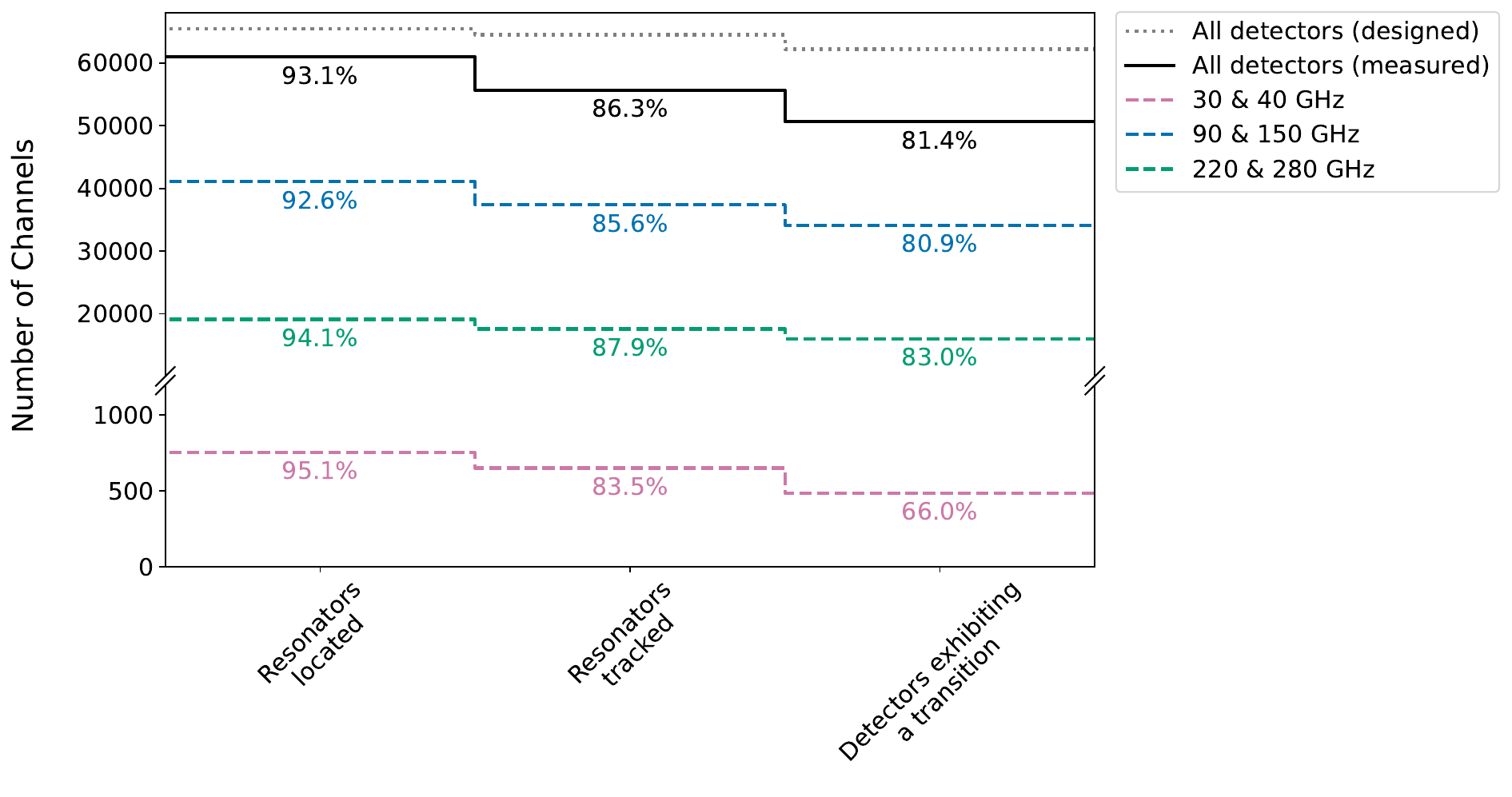}
\caption{Number of channels whose microwave resonators are located during tuning, whose resonances are tracked during flux ramping, and whose detectors exhibit a transition during $I{-}V$ curve acquisition. These counts are split by UFM passband, and they are compared to the total number of channels designed and deployed across the LATR.
\label{fig:mux-measurement}}
\end{figure*}

Figure \ref{fig:mux-measurement} shows each of these three measured values, split by passband and compared to the number of designed channels. As we experienced an issue with the readout chain for one UFM during the analysis time period, which we plan to fix during an upcoming warm-up and opening of the LATR, we omit this module's channels from the measurement and the design count. We find that 93.1\% of microwave resonators and 86.3\% of SQUIDs are identified, with strong consistency across passbands. We also find that 81.4\% of TES detectors exhibit a measured transition, which is in excess of the 70\% and 80\% criteria used for the baseline and goal projections, respectively, presented in \cite{SOScienceGoals-Ade_2019}. The relatively lower fraction of LF TES detectors which exhibit a measured transition may be improved as we continue to optimize their parameters for $I{-}V$ curve acquisition; these detectors with smaller saturation powers are particularly sensitive to heating from the process of driving saturation power down the UFMs' bias lines during this acquisition, therefore affecting the choice of operation bias.

\subsection{Bias line stability}
\label{subsec:bias-stability}

\begin{figure*}[t!]
\centering

\begin{subfigure}[b]{0.99\textwidth}
    \centering
    \includegraphics[width=\textwidth]{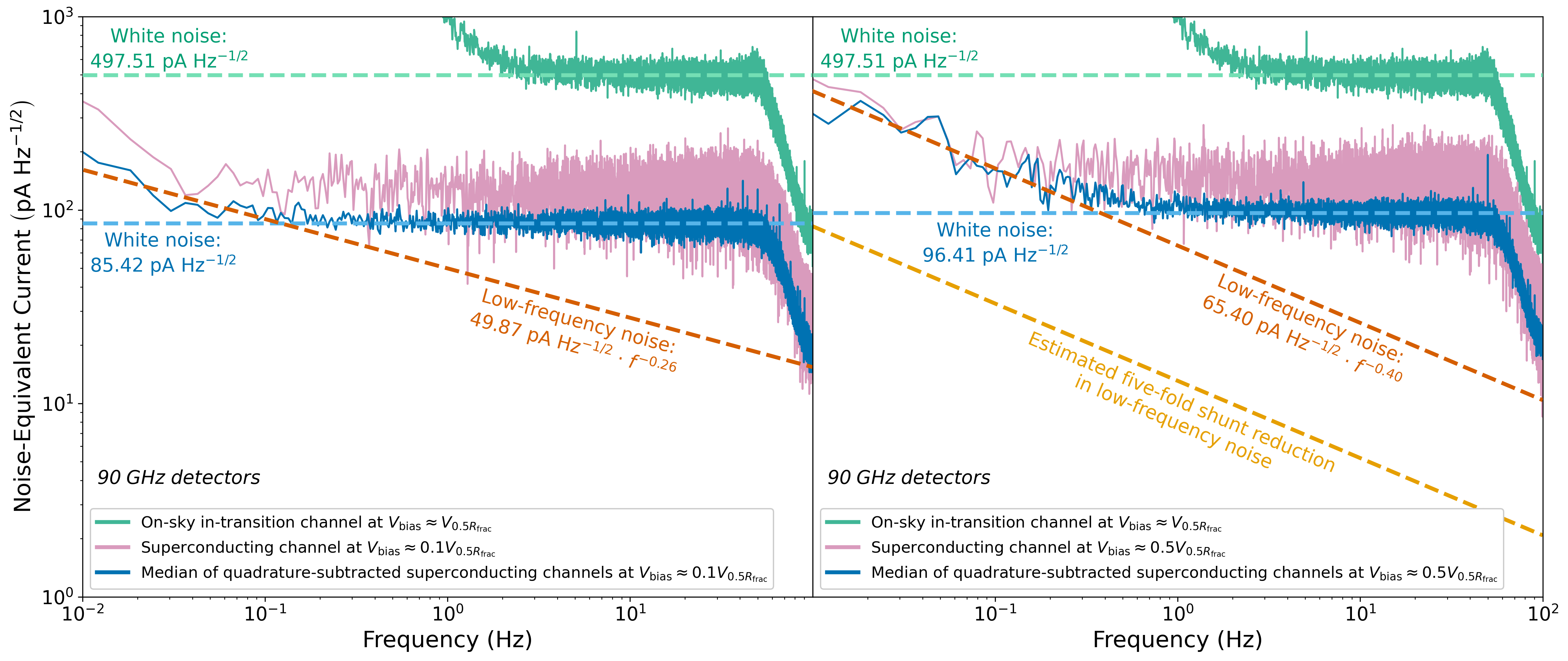}
    \caption{}
    \label{fig:bias-stability-90}
\end{subfigure}
\begin{subfigure}[b]{0.99\textwidth}
    \centering
    \includegraphics[width=\textwidth]{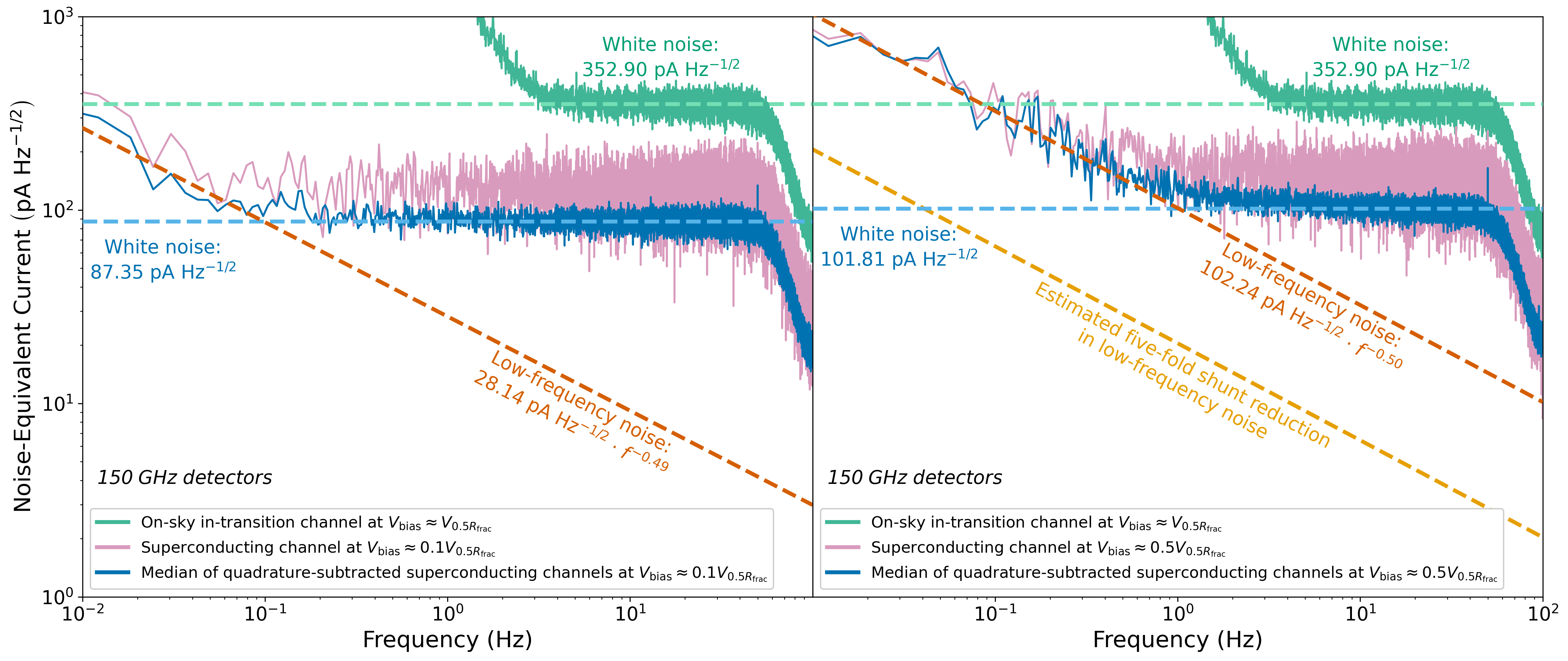}
    \caption{}
    \label{fig:bias-stability-150}
\end{subfigure}
\caption{Noise-equivalent current of superconducting optically-coupled (a) 90\;GHz and (b) 150\;GHz TES detectors on a single bias line each, with an aluminum plate mounted to the end of the OT to remove sky loading. Detectors have (left) one-tenth and (right) one-half of the voltage bias necessary to achieve $R_\text{frac}\approx0.5$ ($V_{0.5R_\text{frac}}$) without first driving them normal. A single representative spectrum is shown in pink. The noise-equivalent current spectrum from a measurement with no voltage on the bias lines is subtracted off in quadrature to isolate the effect of raising the bias line voltage, and the per-frequency-bin median of this spectrum is shown in blue. Fits to the low-frequency and white noise regions are drawn. The right panels show a five-fold reduction in the low-frequency component, as would be conservatively estimated from operating shunted in-transition TES detectors at $R_\text{frac}\approx0.5$. A single representative in-transition channel during a CMB scan is also shown in green. Note that the reduced low-frequency noise is subdominant to this in-transition channel.
\label{fig:bias-stability}}
\end{figure*}

Next, we report on readout system noise performance. The dominant potential source of low-frequency readout noise during observations, which would impact the knee frequency of the optically-coupled channels, may be noise from the TES bias lines. Although significant effort was put into ensuring adequate low-frequency noise performance of SMuRF electronics, this has not yet been empirically verified with a fully-integrated system deployed in the field \citep{SMuRF-Yu_2023}. We measure this noise by biasing the detectors without first driving them normal; by stepping the voltage up directly from 0\;V, the detectors remain superconducting. All of the current from the bias is therefore driven directly through the detectors rather than in parallel with the shunt resistors, and we can measure this via the SQUIDs. We present the noise-equivalent current amplitude spectra of these optically-coupled channels on one UFM with varying fractions of the amount of voltage that would be needed to bias the TES detectors to $R_\text{frac}\approx0.5$ during normal operations ($V_{0.5R_\text{frac}}$). These are computed from two 500-second-long streams of data, one with one-tenth $V_{0.5R_\text{frac}}$, and one with one-half $V_{0.5R_\text{frac}}$. To isolate the effect of the bias, we subtract from each of these, in quadrature, the noise-equivalent current spectrum of a 500-second-long stream of the superconducting TES detectors with no voltage on their bias lines. We perform this measurement with a reflective aluminum plate mounted over the end of the OT in order to remove sky loading and minimize in-band photon noise.

Figure \ref{fig:bias-stability-90} shows the per-frequency-bin median of the quadrature-subtracted noise-equivalent current for 90\;GHz detectors on a single bias line, while Figure \ref{fig:bias-stability-150} shows this spectrum for 150\;GHz detectors on a single bias line of the same UFM. We present measurements of the low-frequency noise, fit between 0.01\;Hz and 0.1\;Hz, and of the white noise, fit between 10\;Hz and 30\;Hz. We observe that the spectral index of the low-frequency noise component rises slightly with the increase in bias voltage, and that there is also an increase in the overall magnitude of this noise. This increase, however, is less than the increase in the bias line voltage. Despite the five-fold increase in bias line voltage between the two measurements, we observe a 2.5x increase in the low-frequency noise at 0.01\;Hz for 90\;GHz detectors, and 3.8x for 150\;GHz detectors. As the detectors have $R_N\approx8\;\text{m}\Omega$ and they are in parallel with shunt resistors of $R_\text{shunt}\approx400\;\mu\Omega$, as shown in Figure \ref{fig:umux-schematic}, this noise would be attenuated by a factor of approximately ten during normal operation at $R_\text{frac}\approx0.5$. Assuming a two-fold increase in low-frequency bias line noise from doubling the bias line voltage from $0.5V_{0.5R_\text{frac}}$---which would be an over-estimate given this measurement---this would predict a five-fold overall decrease in said low-frequency noise during operation. For illustration, we show this five-fold reduction in the right-hand panels in Figure \ref{fig:bias-stability}, which demonstrates a negligible contribution to the low-frequency noise of the representative on-sky in-transition ($V_\text{bias}\approx V_{0.5R_\text{frac}}$) channels shown. Using the white noise levels of these on-sky channels, this reduced low-frequency bias line noise would predict a knee frequency of $\mathcal{O}\left(100\;\mu\text{Hz}\right)$ at 90\;GHz and $\mathcal{O}\left(1\;\text{mHz}\right)$ at 150\;GHz, which is several orders of magnitude below the knee frequency needed to enable the LAT's multipole coverage, as presented in \cite{SOScienceGoals-Ade_2019}. This therefore shows strong performance of the DC signals output by the SMuRF systems' RTMs, whose stability is not significantly contributing to low-frequency noise. Future work will explore the sources and scaling of this noise with bias voltage.

\subsection{Readout white noise}
\label{subsec:readout-noise}

\subsubsection{Open SQUID noise}
\label{subsubsec:open-squids}

\begin{figure*}[t!]
\centering

\begin{subfigure}[b]{0.49\textwidth}
    \centering
    \includegraphics[width=\textwidth]{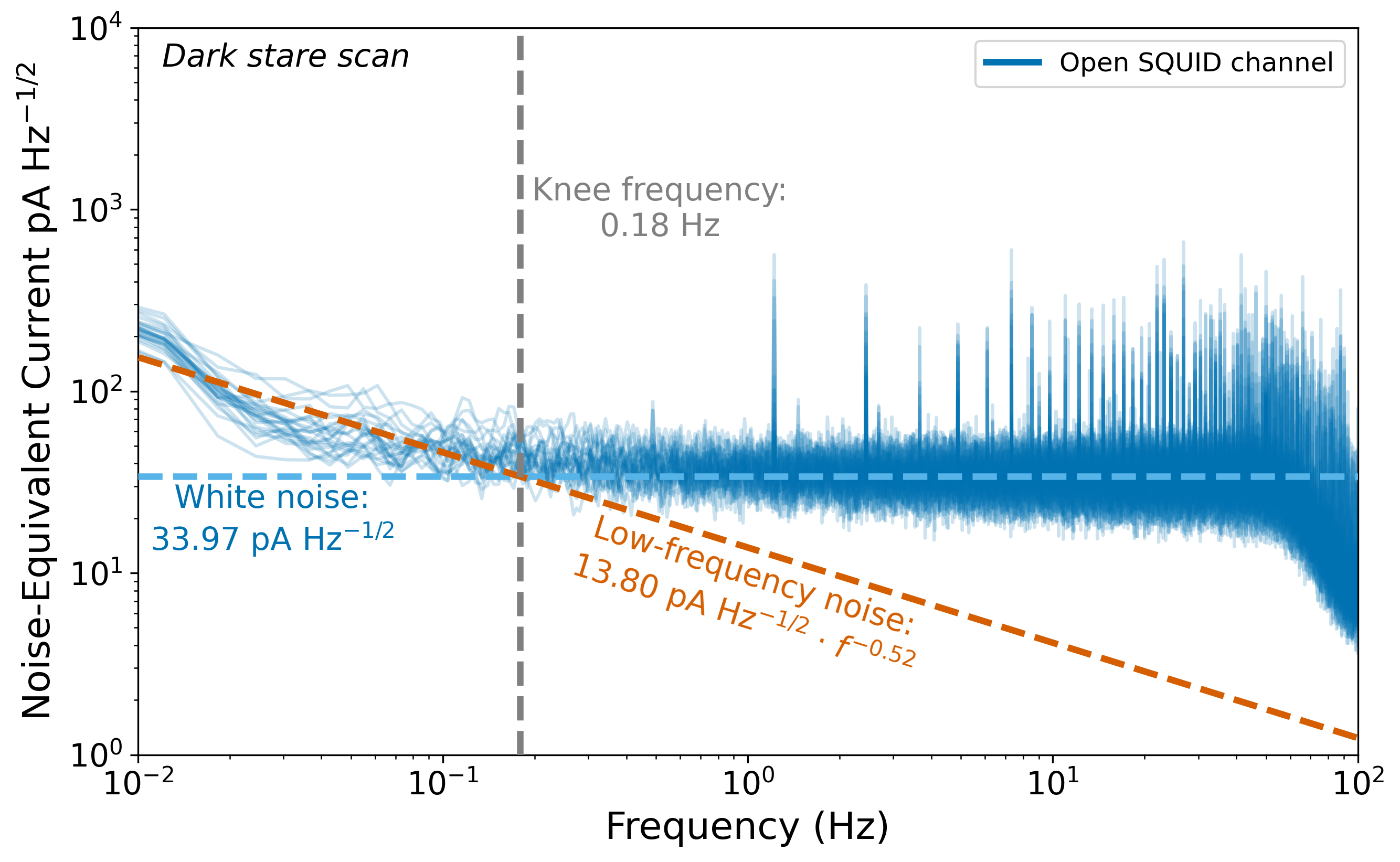}
    \caption{}
    \label{fig:open-squid-stare}
\end{subfigure}
\begin{subfigure}[b]{0.49\textwidth}
    \centering
    \includegraphics[width=\textwidth]{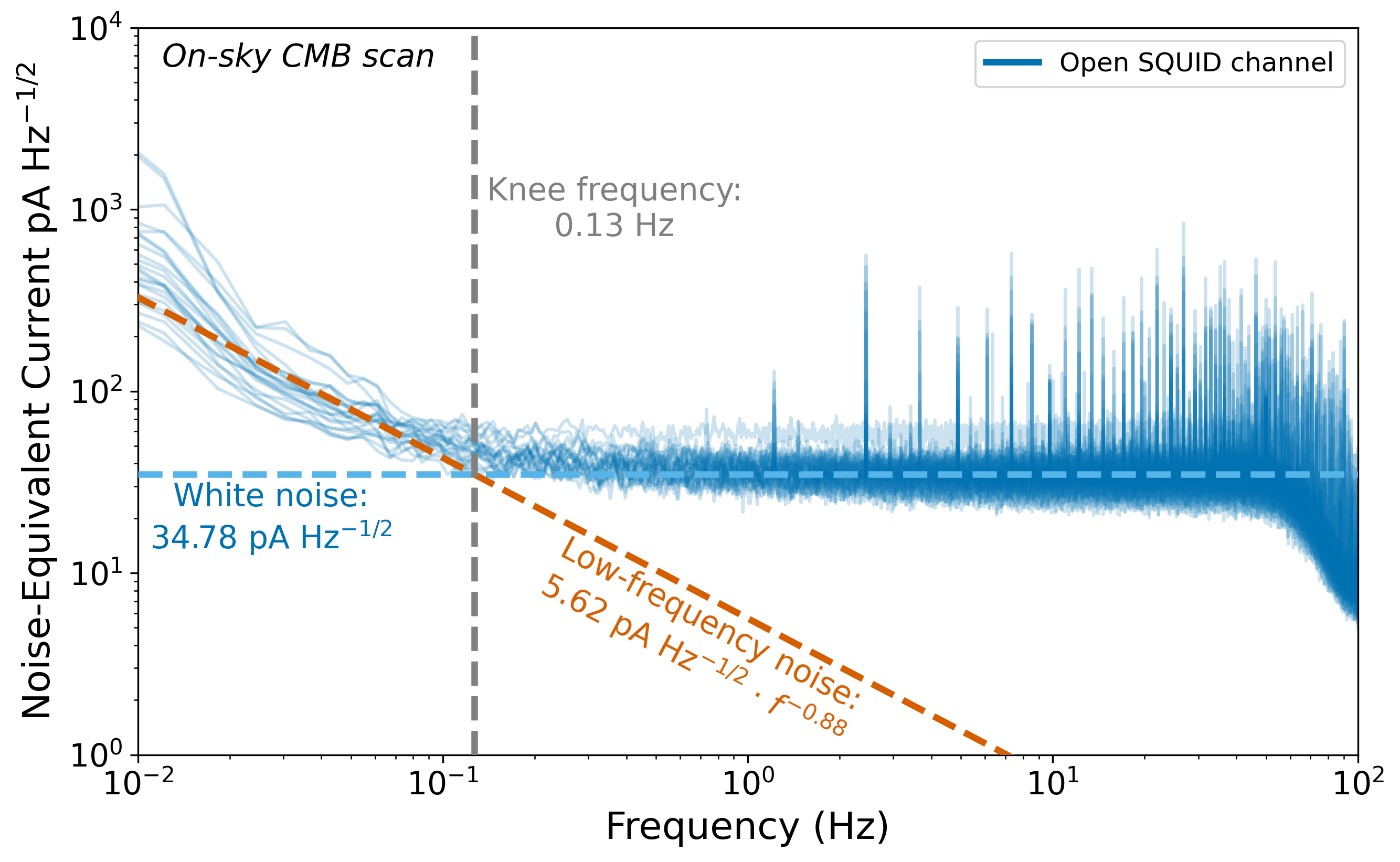}
    \caption{}
    \label{fig:open-squid-scan}
\end{subfigure}
\caption{Noise-equivalent current of the open SQUID channels on a single UFM during (a) a stare with an aluminum cap at the end of the OT and no sky loading, and (b) an on-sky CMB scan. Fits to the low-frequency and white noise regions are drawn, as well as the calculated knee frequency. While we observe a large number of spikes, we find that they are narrow in frequency space and therefore make a small contribution to the signal's variance.
\label{fig:open-squid-nei}}
\end{figure*}

In order to analyze the noise performance of the fully-integrated readout system, including the cryogenic UMMs and the warm SMuRF electronics, we consider the noise measured by the open SQUID channels, which are not coupled to detectors. We identify these channels using the aforementioned resonator matching algorithm \citep{DetMatch-Lashner_2024}. Figure \ref{fig:open-squid-stare} shows the noise-equivalent current of these channels on a single UFM, computed from 20 minutes of streamed data with the telescope stationary, and an aluminum plate mounted to the end of the OT. Figure \ref{fig:open-squid-scan} shows the same statistic, measured during 54 minutes of on-sky CMB observations. We observe white noise performance of roughly 34\;pA\;$\text{Hz}^{-1/2}$ for the dark stare scan, and roughly 35\;pA\;$\text{Hz}^{-1/2}$ for the on-sky CMB scan. These noise levels are well below the white noise from in-transition detectors, as presented in Figure \ref{fig:bias-stability}. We also note that while the magnitude and spectral index of the low-frequency noise change with on-sky scanning, these changes do not appreciably impact the computed knee frequency. One contribution to this low-frequency component may be environmental effects from temperature changes or the telescope's motion during scanning; for a discussion of the former, refer to \cite{PhaseDrift-Satterthwaite_2025}.

\subsubsection{Inferred readout noise}
\label{subsubsec:readout-noise}

We now consider the contribution of readout to the white noise-equivalent current of channels which are coupled to optical TES detectors, $\text{NEI}_\text{readout}$. We optimize this quantity by adjusting RF tone power produced by the SMuRF system to probe the cryogenic resonators, in order to avoid over-driving these resonators. This is the main free parameter of the system that is tunable post-deployment, by digitally adjusting the attenuation from the upconverters and downconverters used to produce the tones within each 500\;MHz SMuRF band. Each of these attenuations has a range of 15\;dB in 0.5\;dB increments. We first determine the minimum amount of combined upconverter and downconverter attenuation which avoids saturating the ADCs and DACs used for probe tone demodulation. We then step through a grid of all possible upconverter and downconverter attenuation combinations which sum to this maximal value and test the white noise level from 30 seconds of data acquired with saturation current driven down the bias lines at each step. As the white noise of normal detectors is the quadrature sum of $\text{NEI}_\text{readout}$ and the Johnson noise-equivalent current $\text{NEI}_\text{Johnson}$, this provides a quick approximation of $\text{NEI}_\text{readout}$ as the $\text{NEI}_\text{Johnson}$ term is a function of TES circuit parameters and temperature.

\begin{figure*}[t!]
\centering
\includegraphics[width=0.7\textwidth]{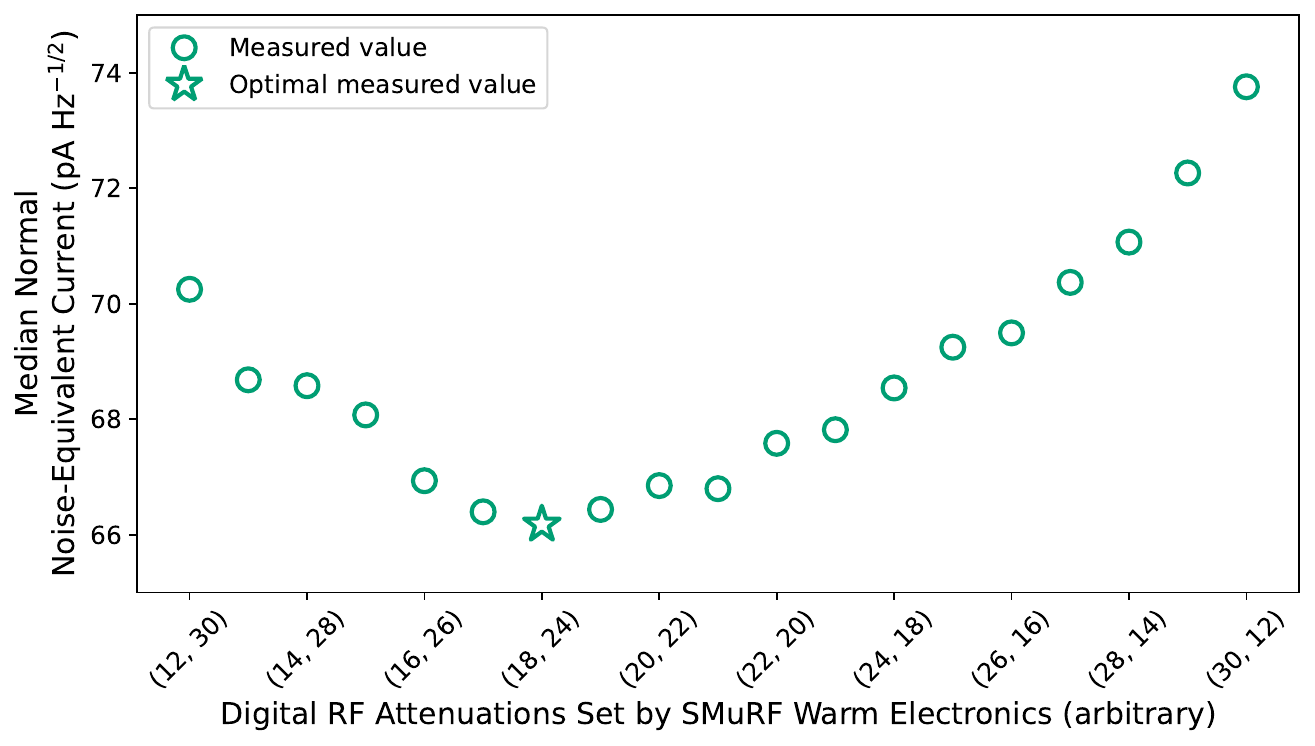}
\caption{Optimization of TES normal noise-equivalent current for 228 channels via changing digital (up-, down-) converter attenuations (arbitrary digital units of 0.5\;dB increments) used in producing and demodulating RF tones. The optimal value is marked with a star.
\label{fig:atten-optimization}}
\end{figure*}

Figure \ref{fig:atten-optimization} shows the result of this optimization for a single 500\;MHz-wide SMuRF band of 228 channels, where the optimal combination of attenuations is marked with a star. We performed this optimization once following the installation of the detector modules and readout system to the LATR, and we use the attenuations which produce the minimum measured white noise for all subsequent data acquisition. We do not expect to repeat this measurement unless we make adjustments to an individual system's RF chain.

We then perform a more detailed measurement of $\text{NEI}_\text{readout}$ by raising the bath temperatures of the UFMs above their detectors' critical temperatures to approximately $200\;\text{mK}$ (with slight variation across the OTs) in order to drive them normal. We compute the $\text{NEI}_\text{Johnson}$ in terms of shunt resistor temperature $T_\text{sh}$, TES temperature $T_\text{TES}$, shunt resistance $R_\text{sh}$, TES resistance $R_\text{TES}$, and the Boltzmann constant $k_B$:

\begin{equation}
    \text{NEI}_\text{Johnson}=\sqrt{\frac{4k_B\left(T_\text{sh} R_\text{sh}+T_\text{TES} R_\text{TES}\right)}{\left(R_\text{sh}+R_\text{TES}\right)^2}}.
\end{equation}

\begin{figure*}[t!]
\centering
\includegraphics[width=0.7\textwidth]{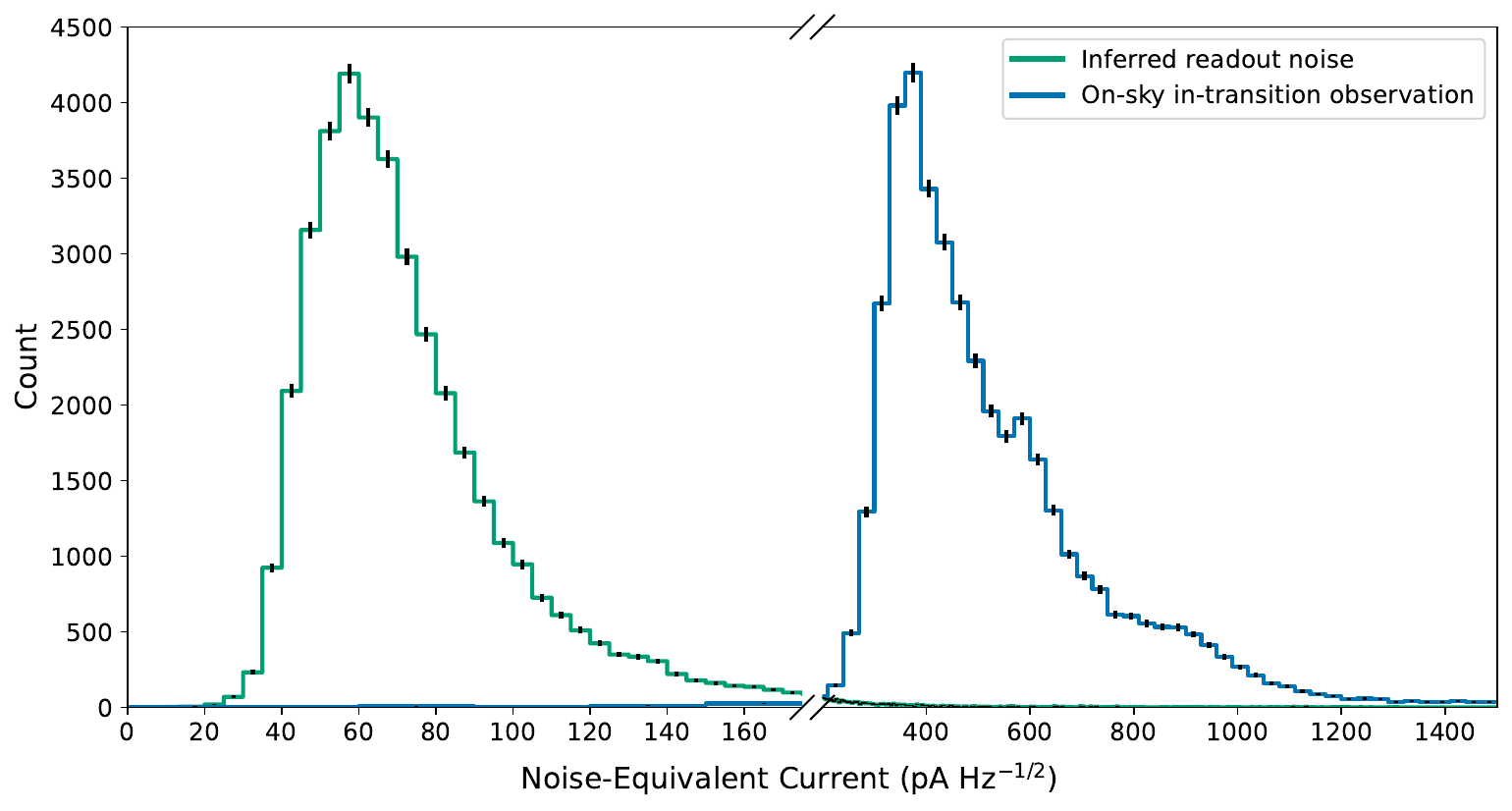}
\caption{Distribution of the readout white noise-equivalent current of optically-coupled TES detectors deployed to the LATR, determined using 15 minutes of noise acquisition with bath temperature of approximately 200\;mK and subtracting off the Johnson noise contribution in quadrature. This is shown on a broken axis with the overall noise-equivalent current of optically-coupled in-transition detectors during a CMB scan. Error bars show one standard deviation Poisson counting uncertainties.
\label{fig:readout-noise}}
\end{figure*}

As the detectors are normal, we approximate $T_\text{bath}=T_\text{sh}=T_\text{TES}$, as measured using a thermometer on each OT's focal plane, and $R_\text{TES}=R_N$, as measured using $I{-}V$ curves. Figure \ref{fig:readout-noise} shows the distribution of $\text{NEI}_\text{Readout}$ for the optically-coupled TES detectors deployed to 37 of the LAT's UFMs, calculated using the median of the amplitude spectral density between 10\;Hz and 30\;Hz during 15 minutes of streamed data and subtracting off the $\text{NEI}_\text{Johnson}$ contribution in quadrature. We plot this distribution on a broken axis alongside the on-sky noise-equivalent current of in-transition optically-coupled detectors. These are measured from a sample on-sky CMB scan for each of the same UFMs, collected between 26 April 2026 and 15 June 2026 during $1.1\;\text{mm}<\text{PWV}/\sin(\text{elevation})<1.5\;\text{mm}$, as we consider $\text{PWV}/\sin(\text{elevation})=1.0\;\text{mm}/\sin(50\degree)\approx1.3\;\text{mm}$ to be nominal observing conditions. We measure a median readout white noise of 68.1\;pA\;$\text{Hz}^{-1/2}$. This is higher than the measured noise of the open SQUID channels presented above; this may be due in part to the fact that resonator quality factor has been shown to decrease upon coupling to detectors \citep{UHFUFM-Dutcher_2025}. Using map noise specifications presented in \cite{SOScienceGoals-Ade_2019}, \cite{UFM-McCarrick_2021} prescribed a target for median readout white noise-equivalent current of 65\;pA\;$\text{Hz}^{-1/2}$; compared to this criterion, this measurement demonstrates that the fully-integrated readout system is performing as expected. Figure \ref{fig:readout-noise} shows that that readout is on the order of a percent-level contribution to overall on-sky in-transition noise.

\section{Conclusion}
\label{sec:conclusion}

We have presented the successful deployment and commissioning of the detector readout system for the Simons Observatory LAT's 63,000 detectors. We described the installation of this system to the LATR, and the various operations that we perform to bias detectors and collect data from the instrument in Section \ref{sec:latr-installation} and Section \ref{sec:det-ops}, respectively. Characterization measurements presented in Section \ref{sec:characterization} established that the system meets key performance benchmarks by identifying superconducting transitions in 81.4\% of detectors, and by facilitating on-sky in-transition detector time stream collection with a minimal low-frequency noise contribution from the TES bias lines and a percent-level readout contribution to the white noise from the fully-integrated system. This represents a key milestone in realizing the LAT's design sensitivity and ability to study a diverse range of probes of the millimeter-wavelength sky. We look forward to presenting studies of the commissioning of the telescope's other systems in future work, which will include a quantification of the detector modules' noise-equivalent temperatures, as well as results from early observations.

\begin{acknowledgments}
This work was funded by grants from the Simons Foundation (MPS-Observatory-00457687, B.K.) and Simons Foundation International (SFI-MPS-Observatory-00007127, B.K.). Simons Observatory operates in the Parque Astron\'omico Atacama in northern Chile under the auspices of the Agencia Nacional de Investigaci\'on y Desarrollo (ANID). We thank the Republic of Chile for hosting Simons Observatory in the northern Atacama, and the local indigenous Licanantay communities, whom we follow in observing and learning from the night sky. This work would not be possible without the Simons Observatory site team based in Chile. We acknowledge the support of REUNA, AmLight, and Parque Astron\'omico in enabling the real-time transfer of data via a fiber-optic connection. This research used resources of the National Energy Research Scientific Computing Center (NERSC), a Department of Energy User Facility (HEP project mp107 2023-2026). This research used the SO:UK Data Centre facility at The University of Manchester, funded by STFC (grant award ST/X006344/1). The work presented in this article was performed on computational resources managed and supported by Princeton Research Computing, a consortium of groups including the Princeton Institute for Computational Science and Engineering (PICSciE) and Research Computing at Princeton University. This work was supported by the U.S. National Science Foundation (Award Number: 2153201). Some of the software developed to produce this analysis can be found publicly available on the Simons Observatory GitHub organization (\href{https://github.com/simonsobs}{https://github.com/simonsobs}). This work was supported in part by the U.S. Department of Energy, Laboratory Directed Research and Development program at SLAC National Accelerator Laboratory, under contract DE-AC02-76SF00515. Work at Argonne National Laboratory was supported by the U.S. Department of Energy, Office of High Energy Physics, under contract DE-AC02-06CH11357. Several figures in this paper were created using the Python packages \texttt{numpy} and \texttt{matplotlib} \citep{numpy,matplotlib}. Anthropic's Claude (Opus 4.7) was used to produce the syntax to render Figure \ref{fig:umux-schematic}; all generated code was reviewed and verified \citep{claude}.
\end{acknowledgments}

\bibliography{references}{}
\bibliographystyle{aasjournalv7}

\end{document}